\documentclass[reprint,superscriptaddress,amsmath,amssymb,floatfix,longbibliography]{revtex4-2}
\usepackage{graphicx}
\usepackage{dcolumn}
\usepackage{float}
\usepackage[dvipsnames]{xcolor}
\usepackage[breaklinks]{hyperref}
\hypersetup{colorlinks=true,linkcolor=blue,citecolor=blue,urlcolor=blue}

\begin{document}


\title{Magnetic field effects on the Kitaev model coupled to environment}

\author{Kiyu Fukui}
\email{k.fukui@aion.t.u-tokyo.ac.jp}
\affiliation{Department of Applied Physics, The University of Tokyo, Bunkyo, Tokyo 113-8656, Japan}

\author{Yasuyuki Kato}
\affiliation{Department of Applied Physics, The University of Tokyo, Bunkyo, Tokyo 113-8656, Japan}
 
\author{Yukitoshi Motome}
\affiliation{Department of Applied Physics, The University of Tokyo, Bunkyo, Tokyo 113-8656, Japan}

\date{\today}

\begin{abstract}
Open quantum systems display unusual phenomena not seen in closed systems, such as new topological phases and unconventional phase transitions.
An interesting example was studied for a quantum spin liquid in the Kitaev model [K. Yang, S. C. Morampudi, and E. J. Bergholtz, Phys. Rev. Lett. \textbf{126}, 077201 (2021)]; an effective non-Hermitian Kitaev model, which incorporates dissipation effects, was shown to give rise to a gapless spin liquid state with exceptional points in the Majorana dispersions. 
Given that an external magnetic field induces a gapped Majorana topological state in the Hermitian case, the exceptional points may bring about intriguing quantum phenomena under a magnetic field. 
Here we investigate the non-Hermitian Kitaev model perturbed by the magnetic field. We show that the exceptional points remain gapless up to a finite critical magnetic field, in stark contrast to the Hermitian case where an infinitesimal field opens a gap. The gapless state is stable over a wide range of the magnetic field for some particular parameter sets, and in special cases, undergoes topological transitions to another gapless state with different winding number around the exceptional points without opening a gap. 
In addition, in the system with edges, we find that the non-Hermitian skin effect is induced by the magnetic field, even for the parameters where the skin effect is absent at zero field. The chirality of edge states is switched through the exceptional points, similarly to the surface Fermi arcs connected by the Weyl points in three-dimensional Weyl semimetals. 
Our results provide a new possible route to stabilize topological gapless quantum spin liquids under the magnetic field in the presence of dissipation.
\end{abstract}

\maketitle
\noindent

\section{Introduction}\label{sec:intro}
The quantum spin liquid (QSL), which is a quantum disordered state in magnets with prominent features such as quantum entanglement and fractional excitations, has been studied as one of the major subjects in modern condensed matter physics from both theoretical and experimental points of view~\cite{Diep, Balents2010, Lacroix, Zhou2017}. Despite the long history of research, well-established examples of the QSL are still limited, and the realization of the QSL in most of the candidate models and materials is still under debate. The celebrated Kitaev model, however, has brought a pivotal breakthrough to this situation, by providing a rare example of QSLs in more than one dimension~\cite{Kitaev2006}.  
Despite strong frustration arising from the bond-dependent anisotropic interactions on a honeycomb lattice, the model is exactly solvable and yields the ground state of a QSL with fractional excitations of itinerant Majorana fermions and localized $\mathbb{Z}_{2}$ fluxes. 
The Majorana fermion excitation constitutes a gapless Dirac-like node at zero magnetic field, but it is gapped out immediately by introducing a magnetic field. 
This gapped state is topologically nontrivial, exhibiting chiral edge modes and non-Abelian anyons under a magnetic field that would be utilized for topological quantum computation~\cite{Freedman2002, Kitaev2003, Kitaev2006, Nayak2008, Ahlbrecht2009}.
Moreover, since the feasibility of the model was proposed for spin-orbit coupled Mott insulators~\cite{Jackeli2009}, a number of intensive searches for the candidate materials have been carried out~\cite{rau2016, Winter2017, Takagi2019, Motome2020a, Motome2020, Trebst2022}, for instance, for Na$_2$IrO$_3$~\cite{Chaloupka2010, Singh2010, Singh2012, Comin2012, Chaloupka2013, Foyevtsova2013, Sohn2013, Katukuri2014, Yamaji2014, HwanChun2015, Winter2016}, $\alpha$-Li$_2$IrO$_3$~\cite{Singh2012, Chaloupka2013, Winter2016}, and $\alpha$-RuCl$_3$~\cite{Plumb2014, Kubota2015, Winter2016, Yadav2016, Sinn2016}.  
In recent years, such exploration has been extended to new family of candidates, such as cobalt compounds~\cite{Liu2018, Sano2018, Liu2020, Kim2022}, iridium ilmenites~\cite{Haraguchi2018, Haraguchi2020, Jang2021}, and $f$-electron compounds~\cite{Jang2019, Xing2020, Jang2020, Ramanathan2021, Daum2021}. In addition to the solid state materials, realization of the model in ultracold atoms~\cite{Duan2003, Micheli2006, Manmana2013, Gorshkov2013, Fukui2022, Sun2023} and Rydberg atoms~\cite{Kalinowski2023, Nishad2023} has also been proposed.

In most studies to date, QSLs have been discussed for closed systems. However, effects of dissipation in open systems can be crucially important for both fundamental physics and applied science perspectives. 
For instance, dissipation is not negligible in the implementation of the Kitaev model by the ultracold atoms. 
In quantum computation devices, the system is essentially open due to the input and readout of quantum information as well as the effect of substrates. Such open quantum systems can display unusual phenomena not seen in closed systems, such as new topological phases~\cite{Zhou2018, Gong2018, Cerjan2019, Kawabata2019} and unconventional critical phenomena~\cite{Horstmann2013, Xiao2019, Matsumoto2020, Hanai2020}. 
While theoretical treatment of open quantum systems coupled to the environment is in general complicated, one of the efficient methods is to study an effective non-Hermitian Hamiltonian which incorporates the effects of dissipation. In recent years, the non-Hermitian systems have been studied intensively, especially from the topological point of view~\cite{Ashida2020, Bergholtz2021, Ding2022, Okuma2022, Zhang2022a, Banerjee2023, Lin2023}.  
For the QSLs in the Kitaev model and its relatives, non-Hermitian effects have been investigated on the honeycomb lattice~\cite{Yang2021, Yang2022}, bilayer honeycomb lattice~\cite{Hwang2023a}, square-octagon lattice~\cite{Rodland2024}, hyperoctagon lattice~\cite{Rodland2024}, two-leg ladder~\cite{Shibata2019}, bilayer square lattice~\cite{Shackleton2024}, and Yao-Lee model~\cite{Yang2023, Mandal2024}.
The non-Hermitian Kitaev model defined on the honeycomb lattice was shown to give rise to an unconventional gapless spin liquid state, in which the Dirac-like node with linear dispersion in the Majorana excitation spectrum in the Hermitian case splits into two exceptional points with square-root dispersions~\cite{Yang2021}. 
In the Hermitian case, the effect of the magnetic field is crucial to make the system gapped and topologically nontrivial as mentioned above, but it has not been clarified yet in the non-Hermitian case, even though the exceptional points may bring about unprecedented quantum phenomena. 

In this paper, we investigate the effect of the magnetic field on the non-Hermitian Kitaev model. Based on the perturbation with respect to the field strength, we show that the exceptional points remain gapless up to a finite critical effective magnetic field, in stark contrast to the Hermitian case where the Dirac-like nodes are gapped out by an infinitesimal magnetic field. We also show that the critical field becomes large beyond the field range where the perturbation theory is justified, and even diverges for some particular parameter sets on the complex planes of the coupling constants. By calculating the winding number of the exceptional points, we find that the system at certain special parameters exhibits an unconventional gapless-gapless topological transition by the magnetic field. 
In addition, in the system with edges, we find that the non-Hermitian skin effect is induced by the magnetic field even for the parameters where it does not occur at zero magnetic field.
We also clarify that the chirality of the edge states is switched at the wave numbers corresponding to the exceptional points, like the Fermi arc surface states connected by the projections of the Weyl points in three-dimensional Weyl semimetals. 
Our results provide a new possible route to stabilize and control topological gapless QSLs under the magnetic field in the presence of dissipation.

The structure of this paper is as follows. In Sec.~\ref{sec:model}, we introduce the non-Hermitian Kitaev model under a magnetic field and briefly review the previous studies for the zero field case. We present our results for the critical effective magnetic field for the gap opening at the exceptional points in Sec.~\ref{subsec:critical_h}, the topological transitions with changes in the winding numbers of the exceptional points in Sec.~\ref{subsec:topo_trans}, and the non-Hermitian skin effect in Sec.~\ref{subsec:skin_eff}. 
In Sec.~\ref{subsec:discuss_topo}, we discuss the topological transition found in this study in comparison with the precedented examples. We also discuss the skin effect in comparison with the surface states in the three-dimensional Hermitian Weyl semimetals in Sec.~\ref{subsec:discuss_Weyl}.
In Sec.~\ref{subsec:discuss_perturbation}, we remark on the effects of the perturbation 
terms not considered in the present calculations. Finally, we summarize our findings in Sec.~\ref{sec:summary}.

\section{Model}\label{sec:model}

\begin{figure}
    \centering
     \includegraphics[width=\columnwidth,clip]{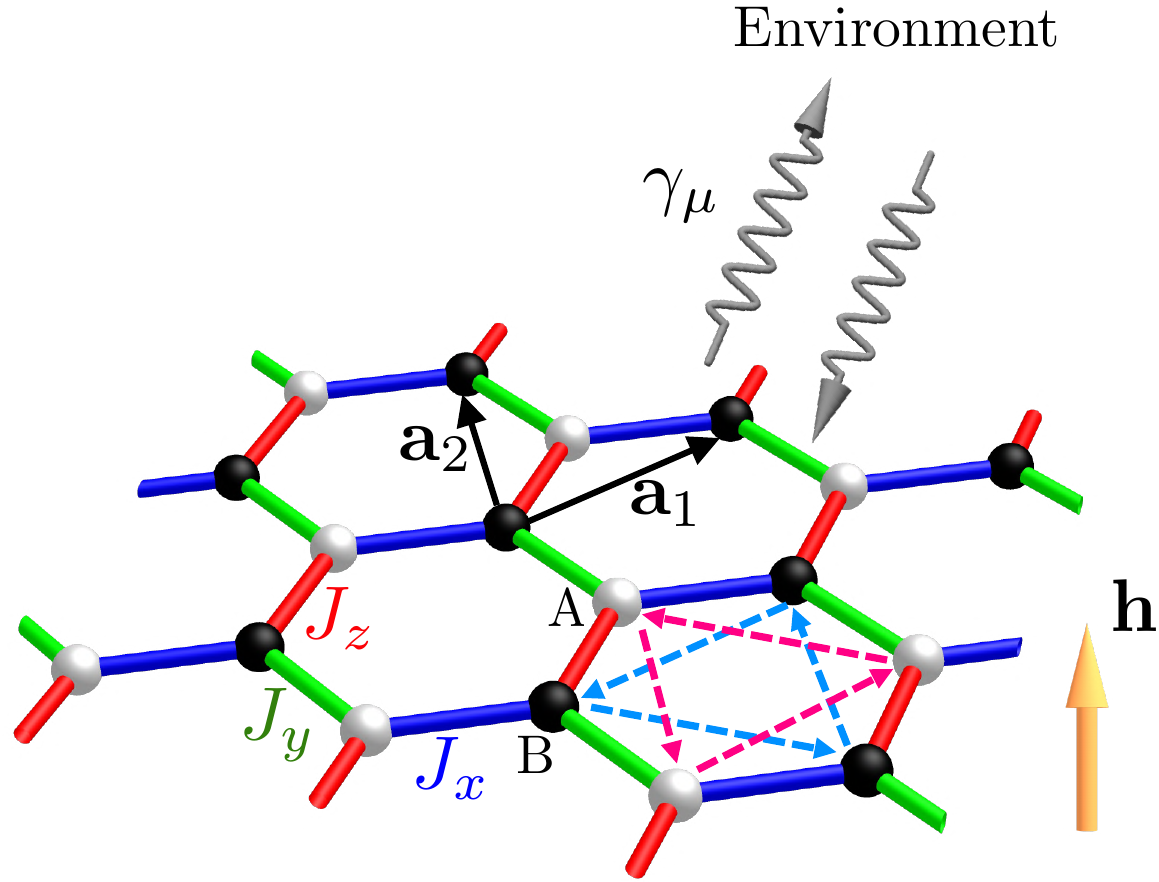}
    \caption{Schematic picture of the Kitaev model coupled to an environment. $J_{\mu}$ and ${\bf h}$ represent the coupling constants of the Kitaev interactions on the $\mu$ bond ($\mu=x$, $y$, and $z$) of the honeycomb lattice and the external magnetic field, respectively; see Eq.~\eqref{eq:hermite}. $\gamma_{\mu}$ represent the coupling to the environment. 
The dashed arrows represent the sublattice-dependent second-neighbor hoppings of Majorana fermions stemming from the magnetic field in the perturbation theory; see Eq.~\eqref{eq:model}. The white (black) spheres represent sites belonging to the A (B) sublattice. $\mathbf{a}_{1}$ and $\mathbf{a}_{2}$ denote the primitive translation vectors.
}
    \label{fig:model}
\end{figure}

We start from the Kitaev model under a uniform magnetic field. The Hamiltonian is given by~\cite{Kitaev2006}
\begin{equation}\label{eq:hermite}
\mathcal{H}_{0}=-\sum_{\langle i, j\rangle_{\mu}}J_{\mu}\sigma^{\mu}_{i}\sigma^{\mu}_{j}-\tilde{h}\sum_{\langle\!\langle i, j, k \rangle\!\rangle_{\mu, \nu, \lambda}}\sigma_{i}^{\mu}\sigma_{j}^{\nu}\sigma_{k}^{\lambda},
\end{equation}
where $\sigma_{i}^{\mu}$ is the $\mu$ component of the Pauli matrix at site $i$ ($\mu=x$, $y$, and $z$). The first term represents the Kitaev interactions on the $\mu$ bond with the coupling constant $J_{\mu}$; the summation of $\langle i, j\rangle_{\mu}$ runs over nearest-neighbor pairs $i$ and $j$ connected by the $\mu$ bond; see Fig.~\ref{fig:model}. 
The second term represents effective three-spin interactions derived by third-order perturbations with respect to the Zeeman coupling; the coupling strength $\tilde{h}$ is proportional to $h_x h_y h_z$ for an external magnetic field ${\bf h} = (h_x,h_y,h_z)$~\cite{Kitaev2006}. The summation of $\langle\!\langle i, j, k\rangle\!\rangle_{\mu, \nu, \lambda}$ runs over neighboring three sites $i$, $j$, and $k$ such that $i$ and $j$ ($j$ and $k$) are connected by $\mu$ ($\lambda$) bond and $\nu$ is chosen as the remaining bond ($\mu\neq\nu\neq\lambda\neq\mu$). The effects of other corrections within the third-order perturbation will be discussed in Sec.~\ref{subsec:discuss_perturbation}.

When the system is coupled to an environment, the time evolution is described by the Lindblad equation for the density matrix $\rho$ given by~\cite{Gorini1975, Lindblad1976}
\begin{equation}
\frac{\mathrm{d}}{\mathrm{d}t}\rho=-\mathrm{i}[\mathcal{H}_{0}, \rho]+\sum_{\langle i, j\rangle_{\mu}}\gamma_{\mu}\left( L_{ij}^{\mu}\rho L_{ij}^{\mu\dagger}-\frac{1}{2}\{L_{ij}^{\mu\dagger}L_{ij}^{\mu}, \rho\}\right),
\label{eq:Lindblad}
\end{equation}
where $L_{ij}^{\mu}$ is a jump operator describing the interactions with environment; we set the reduced Planck constant $\hbar=1$. $\gamma_{\mu}$ is a damping rate and thus it takes a non-negative value. Here, however, we treat it as a real number which also takes a negative value to simplify the mathematical treatment.
In the following, we choose $L_{ij}^{\mu}=\sigma^{\mu}_{i}+\sigma^{\mu}_{j}$, which was introduced in the previous study for zero field~\cite{Yang2021}. 
A similar type of dissipation was considered on the Kitaev model on a two-leg ladder~\cite{Shibata2019}. Since solving this equation directly is complicated and laborious, we deal with an
effective non-Hermitian Hamiltonian given by
\begin{align}
\mathcal{H}_{\mathrm{eff}}&=\mathcal{H}_{0}-\frac{\mathrm{i}}{2}\sum_{\langle i, j\rangle_{\mu}}\gamma_{\mu}L_{ij}^{\mu\dagger} L_{ij}^{\mu} \nonumber\\
&=-\sum_{\langle i, j\rangle_{\mu}}G_{\mu}\sigma^{\mu}_{i}\sigma^{\mu}_{j}-\tilde{h}\sum_{\langle\!\langle i, j, k \rangle\!\rangle_{\mu, \nu, \lambda}}\sigma_{i}^{\mu}\sigma_{j}^{\nu}\sigma_{k}^{\lambda},\label{eq:model_spin}
\end{align}
where $G_{\mu}=J_{\mu}+\mathrm{i}\gamma_{\mu}=\lvert G_{\mu}\rvert\mathrm{e}^{\mathrm{i}\phi_{\mu}}$ is the coupling constant of the Kitaev interactions extended to the complex numbers under the influence of dissipation. 
Note that the first term in Eq.~\eqref{eq:model_spin} is the same as the non-Hermitian Hamiltonian in the previous study at zero field~\cite{Yang2021}.

The dissipative dynamics interspersed with quantum jumps to different states triggered by $L_{ij}^{\mu}\rho L_{ij}^{\mu\dagger}$ in Eq.~\eqref{eq:Lindblad} can be interpreted as the deterministic time evolution by the effective non-Hermitian Hamiltonian in Eq.~\eqref{eq:model_spin}~\cite{Carmichael1993, Plenio1998, Daley2014}. 
The feasibility of such a time domain was also discussed in the previous study~\cite{Yang2021}.
This non-Hermitian model is exactly solvable through the Majorana fermion representation introduced in the Kitaev's seminal paper~\cite{Kitaev2006}, in which spins are represented by four types of Majorana fermions ($c_{i}$, $b_{i}^{x}$, $b_{i}^{y}$, $b_{i}^{z}$) as $\sigma^{\mu}_{i}=\mathrm{i}b_{i}^{\mu}c_{i}$. 
In this representation, the spin degree of freedom is fractionalized into itinerant Majorana fermions $c_{i}$ and the localized $\mathbb{Z}_{2}$ fluxes defined as $W_{p}=\prod_{\langle i, j\rangle_{\mu}\in p} u_{ij}^{\mu}$ on each hexagonal plaquette labeled $p$ of the honeycomb lattice; $u_{ij}^{\mu}=\mathrm{i}b_{i}^{\mu}b_{j}^{\mu}$ is a gauge variable taking $\pm1$ on the $\mu$ bond connecting site $i$ and $j$ belonging to the A and B sublattice, respectively. The latter $\mathbb{Z}_2$ fluxes are conserved quantities even in the non-Hermitian case, which enables us to group the eigenstates into different sectors of the flux configuration as in the Hermitian case. Although Lieb's theorem~\cite{Lieb1994} guarantees that the ground state is flux free ($W_{p}=1$ for all plaquettes) in the Hermitian case, there is no longer a mathematical theorem for the non-Hermitian case that tells us the flux configuration in the ground state, which is defined as a state with the lowest energy and the longest lifetime.
 In the previous study at zero magnetic field, it is numerically confirmed that the flux-free state has the lowest energy and the longest lifetime for some parameters~\cite{Yang2021}. We assume that this holds for the present case perturbed by a weak magnetic field. Then, by choosing all the gauge variables as $u_{ij}^{\mu}=1$, the Hamiltonian in Eq.~\eqref{eq:model_spin} is written by a quadratic form of the Majorana fermion $c_{i}$ as
\begin{equation}\label{eq:model}
\mathcal{H}_{\mathrm{eff}}=\mathrm{i}\sum_{\langle i,j\rangle_{\mu}}G_{\mu} c_{i}c_{j}-\mathrm{i}\tilde{h}\sum_{\{i, j\}}c_{i}c_{j},
\end{equation}
where the second term denotes sublattice-dependent second-neighbor hoppings of the Majorana fermions; the summation $\{i, j\}$ runs over second-neighbor pairs shown by the dashed arrows from $j$ to $i$ in Fig.~\ref{fig:model}. 

In Secs.~\ref{subsec:critical_h} and \ref{subsec:topo_trans}, we study this model under the periodic boundary condition (PBC). In this case, we consider the Fourier transform of Eq.~\eqref{eq:model} summarized into
\begin{equation}\label{eq:model_k}
\mathcal{H}_{\mathrm{eff}}
=\sideset{}{^{'}}\sum_{\mathbf{k}}
\begin{pmatrix}
c^{\dagger}_{\mathbf{k}, \mathrm{A}} && c^{\dagger}_{\mathbf{k}, \mathrm{B}}
\end{pmatrix}
H(\mathbf{k})
\begin{pmatrix}
c_{\mathbf{k}, \mathrm{A}} \\ c_{\mathbf{k}, \mathrm{B}}
\end{pmatrix},
\end{equation}
where
\begin{align}
H(\mathbf{k})
=\begin{pmatrix}
\Delta(\mathbf{k}) && \mathrm{i}\varepsilon(\mathbf{k}) \\
-\mathrm{i}\varepsilon(-\mathbf{k}) && -\Delta(\mathbf{k})
\end{pmatrix}, \label{eq:H_k}
\end{align}
with
\begin{align}
\varepsilon(\mathbf{k})&=2(G_{x}\mathrm{e}^{\mathrm{i}\mathbf{k}\cdot\mathbf{a}_{1}}+G_{y}\mathrm{e}^{\mathrm{i}\mathbf{k}\cdot\mathbf{a}_{2}}+G_{z}) \label{eq:epsilon_k},\\
\Delta(\mathbf{k})&=4\tilde{h}[-\sin(\mathbf{k}\cdot\mathbf{a}_{1})+\sin(\mathbf{k}\cdot\mathbf{a}_{2})+\sin\{\mathbf{k}\cdot(\mathbf{a}_{1}-\mathbf{a}_{2})\}].
\end{align}
Here, $\mathbf{a}_{1}=(1/2,\ \sqrt{3}/2)$ and $\mathbf{a}_{2}=(-1/2,\ \sqrt{3}/2)$ denote the primitive translation vectors as shown in Fig.~\ref{fig:model}, and the Fourier transformations of the Majorana fermion operators are defined as
\begin{align}
c^{\dagger}_{\mathbf{k}, \mathrm{A}(\mathrm{B})}&=\frac{1}{\sqrt{2N}}\sum_{\mathbf{r}}\mathrm{e}^{\mathrm{i}\mathbf{k}\cdot\mathbf{r}}c^{\dagger}_{\mathbf{r}, \mathrm{A}(\mathrm{B})},\\
c_{\mathbf{k}, \mathrm{A}(\mathrm{B})}&=\frac{1}{\sqrt{2N}}\sum_{\mathbf{r}}\mathrm{e}^{-\mathrm{i}\mathbf{k}\cdot\mathbf{r}}c_{\mathbf{r}, \mathrm{A}(\mathrm{B})},
\end{align}
where $\mathbf{r}$ represents the position of the unit cell to which the Majorana fermion belongs, and $N$ is the number of the unit cell. The primed summation in Eq.~\eqref{eq:model_k} runs over a half of the first Brillouin zone. 
The Hamiltonian is easily diagonalized and the eigenvalues are obtained as 
\begin{equation}\label{eq:eigenval}
E_{\pm}(\mathbf{k})=\pm\sqrt{\varepsilon(\mathbf{k})\varepsilon(-\mathbf{k})+\Delta(\mathbf{k})^{2}}.
\end{equation}
Note that $E_{\pm}(\mathbf{k})\in\mathbb{C}$ because $\varepsilon(-\mathbf{k})\neq\varepsilon^{*}(\mathbf{k})$, unlike the Hermitian case where the eigenvalues are always real. The square root of a complex function is defined as its principal value.

In the absence of the magnetic field ($\tilde{h}=0$), it was shown that the model has two kinds of phases in the ground state~\cite{Yang2021}: a gapped quantum spin liquid phase (A phase) and a gapless quantum spin liquid phase (B phase).
The latter appears in the region where 
\begin{equation}\label{eq:gapless}
\lvert G_{x}\rvert \leq \lvert G_{y}\rvert + \lvert G_{z}\rvert,\ \lvert G_{y}\rvert \leq \lvert G_{z}\rvert + \lvert G_{x}\rvert,\ \lvert G_{z}\rvert \leq \lvert G_{x}\rvert + \lvert G_{y}\rvert.
\end{equation}
Within the gapless phase, a Dirac-like node with the linear dispersion in the Hermitian case is split into two exceptional points with the square-root dispersions.
We will investigate the effects of the magnetic field in this phase in the following sections. 

We also study the model under the open boundary condition (OBC) in one direction while keeping the PBC in the other direction.
In the non-Hermitian system with edges, the so-called skin effect can occur by accumulating the eigenfunctions near the edges~\cite{Kunst2018, MartinezAlvarez2018, Xiong2018, Yao2018}. 
The condition for the non-Hermitian skin effect was studied for the model in Eq.~\eqref{eq:model} at zero field~\cite{Yang2021}. The details of the setup and calculations will be shown in the beginning of Sec.~\ref{subsec:skin_eff}. 
We will show that the skin effect appears in a richer form in the presence of the magnetic field.

\section{Result}\label{sec:result}
\subsection{Critical effective magnetic field}\label{subsec:critical_h}

\begin{figure}
    \centering
     \includegraphics[width=\columnwidth,clip]{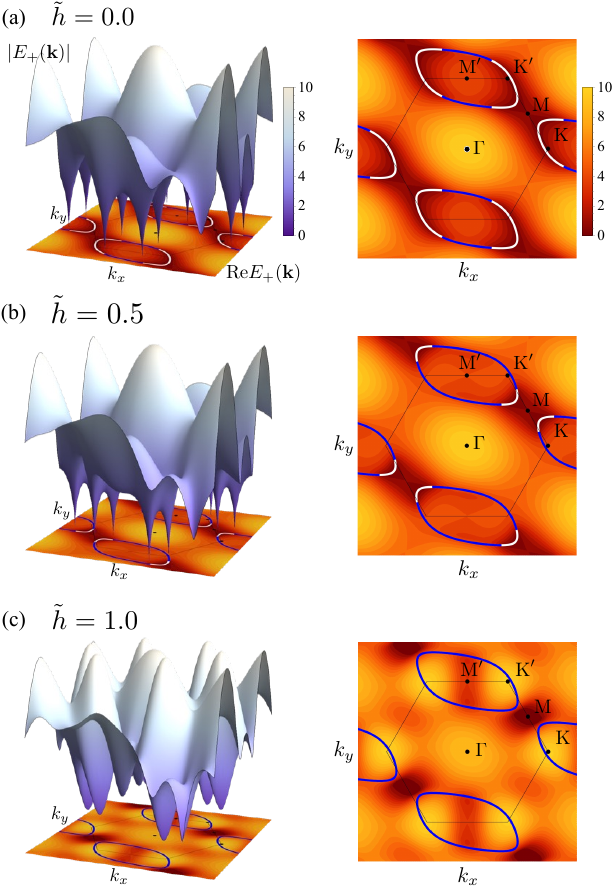}
    \caption{Majorana excitation spectra of the model in Eq.~\eqref{eq:model_k} with $G_{x}=2.0$, $G_{y}=1.0$, and $G_{z}=2.5\mathrm{e}^{\mathrm{i}\frac{\pi}{3}}$ under the effective magnetic field of (a) $\tilde{h}=0.0$, (b) $\tilde{h}=0.5$, and (c) $\tilde{h}=1.0$.
    The left panels show the absolute value of the complex energy eigenvalue, $\lvert E_{+}(\mathbf{k})\rvert$,
together with the real part $\mathrm{Re}E_{+}(\mathbf{k})$ on the bottom plane [see Eq.~\eqref{eq:eigenval}]. The right panels show $\mathrm{Re}E_{+}(\mathbf{k})$. The white and blue lines show the real and imaginary Fermi arcs, which are defined by $\mathrm{Re}E_{+}(\mathbf{k})=0$ and $\mathrm{Im}E_{+}(\mathbf{k})=0$, respectively. 
    The black hexagon indicates the first Brillouin zone, and $\Gamma$, $\mathrm{K}$, $\mathrm{K}'$, $\mathrm{M}$, and $\mathrm{M}'$ denote the high symmetry points.
}
    \label{fig:dispersion1}
\end{figure}

\begin{figure}
    \centering
     \includegraphics[width=0.8\columnwidth,clip]{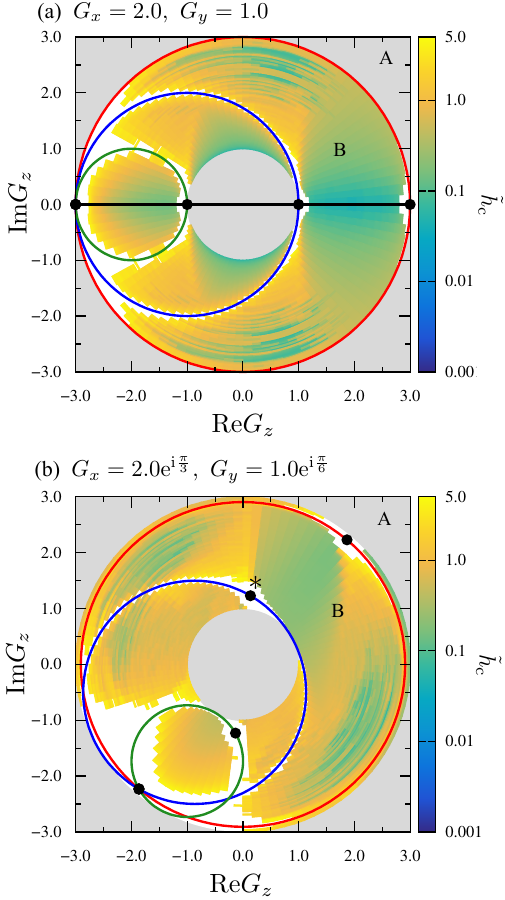}
    \caption{Contour plots of the critical effective magnetic field $\tilde{h}_{\mathrm{c}}$ on the complex plane of $G_z$ at (a) $G_{x}=2.0$ and $G_{y}=1.0$ and (b) $G_{x}=2.0\mathrm{e}^{\mathrm{i}\frac{\pi}{3}}$ and $G_{y}=1.0\mathrm{e}^{\mathrm{i}\frac{\pi}{6}}$. The gray and white regions indicate the gapped A phase at zero magnetic field and the regions where $\tilde{h}_{\mathrm{c}}>5.0$, respectively. The blue, green, and red rings show the analytical results in Eq.~\eqref{eq:ring_xyz}. The black dots indicate the parameters at which $\varepsilon(\mathbf{k})=\varepsilon(-\mathbf{k})=0$; see Eqs.~\eqref{eq:dots_xyz} and \eqref{eq:dot_xyz}.
The black dot with an asterisk in (b) denotes the parameters used in Fig.~\ref{fig:topo_trans}. 
The black horizontal line at ${\rm Im}G_z=0$ in (a) indicates the Hermitian case. 
}
    \label{fig:hc}
\end{figure}

First, we study how the applied magnetic field modifies the Majorana excitation spectrum. 
Figure~\ref{fig:dispersion1} displays the typical spectra by plotting $\lvert E_{+}(\mathbf{k})\rvert$ and $\mathrm{Re}E_{+}(\mathbf{k})$ for three values of the effective magnetic field. Here we choose $G_{x}=2.0$, $G_{y}=1.0$, and $G_{z}=2.5\mathrm{e}^{\mathrm{i}\frac{\pi}{3}}$. These parameters satisfy the conditions in Eq.~\eqref{eq:gapless}, i.e., the system is in the gapless B phase at zero field, as shown in Fig.~\ref{fig:dispersion1}(a); the Majorana fermion dispersion shows exceptional points with square-root ${\bf k}$ dependence that appear at the boundaries between the real and imaginary Fermi arcs defined as $\mathrm{Re} E_{+}(\mathbf{k})=0$ and $\mathrm{Im} E_{+}(\mathbf{k})=0$, shown by the white and blue lines, respectively.
When $\tilde{h}$ is introduced, the system remains gapless at the exceptional points, as shown in Fig.~\ref{fig:dispersion1}(b). 
While increasing $\tilde{h}$, the exceptional points move to shrink the real Fermi arcs, and finally, annihilate in pairs at a critical effective magnetic field $\tilde{h}_{\mathrm{c}} \simeq 0.78$. 
For $\tilde{h} > \tilde{h}_{\mathrm{c}}$, the Majorana spectrum is gapped, as shown in Fig.~\ref{fig:dispersion1}(c). See Appendix~\ref{app:comp_spec} for the detailed changes in the complex energy spectra.

Thus, in the non-Hermitian case, the exceptional points are gapped out at a nonzero effective magnetic field. 
This is in stark contrast to the Hermitian case where the Dirac-like linear dispersion is gapped out by an infinitesimal effective magnetic field~\cite{Kitaev2006}. The qualitative difference stems from the different number of conditions for the gapless states between the Hermitian and non-Hermitian cases~\cite{Berry2004, Bergholtz2021}: In the Hermitian case, three conditions are required in the $2\times2$ Bloch Hamiltonian matrix for gap closing, but they are reduced to two in the non-Hermitian case, thus stabilizing the gapless state more easily. 

The value of the critical field $\tilde{h}_{\mathrm{c}}$ depends on the parameters $G_\mu$. 
We plot $\tilde{h}_{\mathrm{c}}$ for two sets of $G_x$ and $G_y$ on the complex $G_{z}$ plane in Fig.~\ref{fig:hc}. 
Here, we numerically estimate $\tilde{h}_{\mathrm{c}}$ by the value for which the minimum of $\lvert E_{+}(\mathbf{k})\rvert^{4}$ exceeds a threshold of $5.0\times 10^{-4}$~\cite{Note1} while increasing $\tilde{h}$ from $0.0$ to $5.0$ in $0.001$ increments; we
calculate $E_{+}(\mathbf{k})$ on $500\times500$ $\mathbf{k}$-point mesh.
The gray regions indicate the A phase, i.e., in these regions the Majorana spectrum is already gapped  
at $\tilde{h}=0$. The white regions indicate the parameter regions where $\tilde{h}_{\mathrm{c}}>5.0$.
We find that $\tilde{h}_{\mathrm{c}}$ is nonzero in all regions where the system is in the gapless B phase at zero field, and exhibits complicated dependences on the coupling constants $G_\mu$, except for the black line in Fig.~\ref{fig:hc}(a) corresponding to the Hermitian case. 
 
In Fig.~\ref{fig:hc}, the white regions where $\tilde{h}_{\mathrm{c}}$ becomes large appear in specific regions on the $G_z$ plane. 
To understand the origin of these behaviors, we analyze the $G_{\mu}$ dependence of the critical field $\tilde{h}_c$, and find that $\tilde{h}_c$ diverges, within the perturbation theory, for the parameters
\begin{align}
\lvert G_{x}\rvert=\lvert G_{y}+G_{z}\rvert,\ 
\lvert G_{y}\rvert =\lvert G_{z}+G_{x}\rvert,\ 
\lvert G_{z}\rvert =\lvert G_{x}+G_{y}\rvert. 
\label{eq:ring_xyz}
\end{align}
See Appendix~\ref{app:ring} for the derivation. 
These divergence conditions are shown by the blue, green, and red rings in Fig.~\ref{fig:hc}; the white regions appear along these three rings.
Furthermore, assuming that $\varepsilon(\mathbf{k})=0$ and $\varepsilon(-\mathbf{k})=0$ are satisfied simultaneously at the same $\mathbf{k}$, we obtain four more special solutions from Eq.~\eqref{eq:ring_xyz}: 
\begin{align}
G_{x}&=G_{y}+G_{z},\ 
G_{y}=G_{z}+G_{x},\ 
G_{z}=G_{x}+G_{y}, 
\label{eq:dots_xyz}\\
G_{z}&=-G_{x}-G_{y}. \label{eq:dot_xyz}
\end{align}
We also show these solutions by the black dots in Fig.~\ref{fig:hc}. Equation~\eqref{eq:dots_xyz} correspond to the three black dots on the blue, green, and red ring, while Eq.~\eqref{eq:dot_xyz} corresponds to the dot at the intersection of the three rings.
We note that the white regions along the rings become wide near these four dots, especially the last one for Eq.~\eqref{eq:dot_xyz}, indicating strong divergence around these parameters.

\subsection{Topological transitions by the effective magnetic field}\label{subsec:topo_trans}

\begin{figure*}
    \centering
     \includegraphics[width=\linewidth,clip]{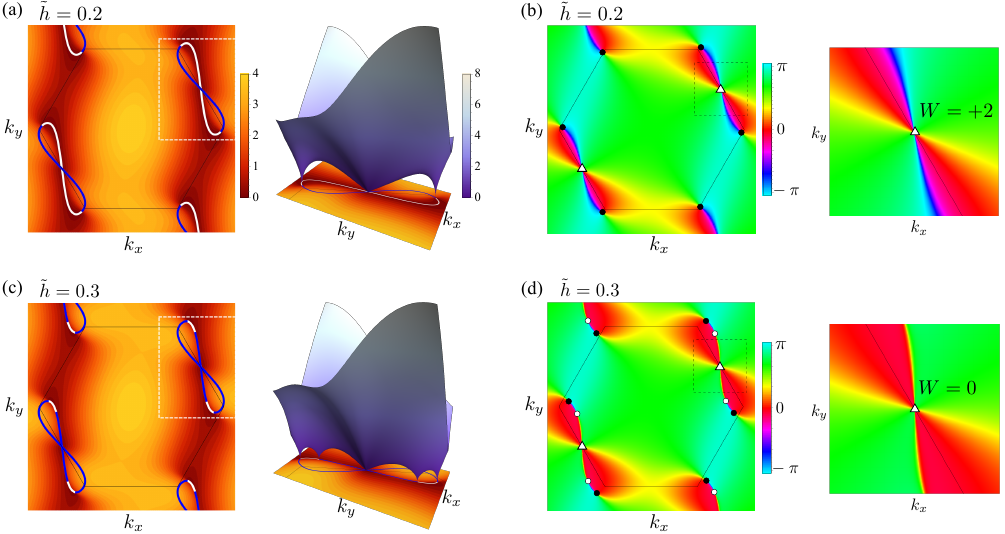}
    \caption{Topological transitions caused by increasing $\tilde{h}$.
In (a) and (c), the left panels show $\mathrm{Re}E_{+}(\mathbf{k})$, and the right panels show $\lvert E_{+}(\mathbf{k})\rvert$ with $\mathrm{Re}E_{+}(\mathbf{k})$ on the bottom plane within the regions indicated by the white dashed rectangles in the left panels. In (b) and (d), the left panels show $\mathrm{Arg}[\mathrm{det}H(\mathbf{k})]$, and the right panels show enlarged views of the regions indicated by the black dashed squares in the left panels. The data are plotted for the parameters $G_\mu$ indicated by the black dot with an asterisk in Fig.~\ref{fig:hc}(b) at $\tilde{h}=0.2$ in (a) and (b), and $\tilde{h}=0.3$ in (c) and (d).
    The notations in (a) and (c) are common to those in Fig.~\ref{fig:dispersion1}. The black dots in (b) and (d) indicate the exceptional points, and the white ones in (d) indicate those generated in pairs by the topological transition.
The white triangles indicate the exceptional points that change the associated winding numbers $W$ in Eq.~\eqref{eq:winding} through the topological transition.
}
    \label{fig:topo_trans}
\end{figure*}

\begin{figure}
    \centering
     \includegraphics[width=\columnwidth,clip]{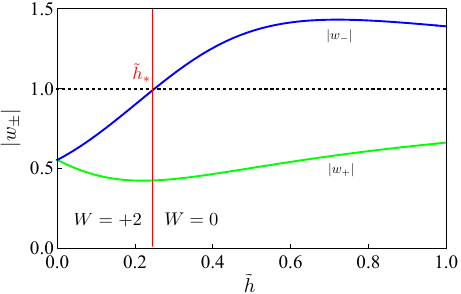}
    \caption{$\tilde{h}$ dependences of the absolute values of poles $\lvert w_{+}\rvert$ (green) and $\lvert w_{-}\rvert$ (blue) in Eq.~\eqref{eq:winding_wpm} for the same $G_\mu$ as Fig.~\ref{fig:topo_trans}. The red vertical line indicates the effective magnetic field $\tilde{h}_{*}$ at which the topological transition occurs with the change in the winding number $W$ from $2$ to $0$ by $|w_{-}|$ exceeding $1$.
}
    \label{fig:poles}
\end{figure}

The exceptional points correspond to the zeros of the complex functions $E_{\pm}(\mathbf{k})$. One can define the winding number around each of them as~\cite{Kawabata2019b}
\begin{equation}\label{eq:winding}
W=\oint_{C}\frac{\mathrm{d}\mathbf{k}}{2\pi\mathrm{i}}
\nabla_{\mathbf{k}}\ln\det[H(\mathbf{k})-E(\mathbf{k}_{\mathrm{EP}})],
\end{equation}
where $H(\mathbf{k})$ and $E(\mathbf{k}_{\mathrm{EP}})$ represent the Bloch Hamiltonian with the wave vector $\mathbf{k}$ 
in Eq.~\eqref{eq:H_k}
and the energy eigenvalue at an exceptional point $\mathbf{k}=\mathbf{k}_{\mathrm{EP}}$, respectively; the contour $C$ is taken to enclose an exceptional point counterclockwise. 
Performing the integration, we obtain $W=\pm1$ for the exceptional points with the square-root dispersion in Figs.~\ref{fig:dispersion1}(a) and \ref{fig:dispersion1}(b). 
Similarly, the winding number takes $\pm1$ up to $\tilde{h}_{\mathrm{c}}$ for generic parameters included in the B phase, whereas it is no longer well defined for $\tilde{h}>\tilde{h}_{\mathrm{c}}$ since the exceptional points are gapped out. Details of the calculations of the winding number are described in Appendix~\ref{app:winding}.

For the special parameters satisfying one of Eqs.~\eqref{eq:dots_xyz} and \eqref{eq:dot_xyz}, however, we find a topological transition associated with the change of $W$ within the gapless phase. 
In these cases, the system exhibits two types of exceptional points at zero field: one with square-root dispersion as for the other parameters, and the other with linear dispersion satisfying $\varepsilon(\mathbf{k}_{\mathrm{EP}})=\varepsilon(-\mathbf{k}_{\mathrm{EP}})=0$. 
In the latter, the linear dispersion appears as $\varepsilon(\mathbf{k})$ and $\varepsilon(-\mathbf{k})$ become zero simultaneously at the same $\mathbf{k}=\mathbf{k}_{\mathrm{EP}}$, and hence, these exceptional points have the winding number $W=\pm2$ each at zero field. Interestingly, we find that the winding number changes from $W=\pm2$ to $W=0$ while increasing $\tilde{h}$, indicating the topological transition. 

We demonstrate this in Fig.~\ref{fig:topo_trans} for the parameters indicated by the black dot with an asterisk in Fig.~\ref{fig:hc}(b). 
In this case, below the critical field $\tilde{h}_*$ (in this specific case, $\tilde{h}_*=0.25$; see Appendix~\ref{app:h_star}), we find the exceptional point with linear dispersion at the crossing point of the real and imaginary Fermi arcs as shown in Fig.~\ref{fig:topo_trans}(a), in addition to a pair of the exceptional points with square-root dispersion. 
The winding number for the former takes $W=+2$, while that for the latter is $W=\pm 1$, as shown in the plot of the complex phase of the Bloch Hamiltonian in Fig.~\ref{fig:topo_trans}(b). 
With the increase of $\tilde{h}$ above $\tilde{h}_*$, the former turns into $W=0$ through the topological transition. 
This is caused by pair creations of additional exceptional points indicated by the white dots in the left panel of Fig.~\ref{fig:topo_trans}(d), leading to the splitting of the real Fermi arcs indicated by the white lines in Figs.~\ref{fig:topo_trans}(a) and \ref{fig:topo_trans}(c). 
We note that the exceptional point with $W=0$ retains linear dispersion, as shown in Fig.~\ref{fig:topo_trans}(c).

The changes of the winding number at the topological transition can be understood from the analysis of the poles in the integrand of Eq.~\eqref{eq:winding}. 
By taking the contour $C$ as a sufficiently small circle of radius $r$ around the exceptional point with parametrizing $\mathbf{k}$ as $\mathbf{k}=\mathbf{k}_{\mathrm{EP}}+r(\cos\theta,\ \sin\theta)$ ($0\leq\theta<2\pi$) and transforming the integral variable as $z=\mathrm{e}^{2\mathrm{i}\theta}$, we obtain a formula for the winding number as
\begin{align}\label{eq:winding_wpm}
W=\oint_{\lvert z\rvert=1}\frac{\mathrm{d}z}{2\pi\mathrm{i}}\frac{2}{z}
\frac{z^{2}-w_{+}w_{-}}{(z-w_{+})(z-w_{-})},
\end{align}
where $r$ appears in both the denominator and numerator and thus they cancel each other out.
See Appendix~\ref{app:winding} for the detailed calculations. The integrand has three poles: $z=0$, $w_{+}$, and $w_{-}$ with residues $-2$, $+2$, and $+2$, respectively. The pole at $z=0$ is always in the unit circle, while the positions of the other two $z=w_\pm$ depend on $\tilde{h}$ as well as $G_\mu$. When both are inside the unit circle at $\tilde{h}=0$, one of them moves outside the unit circle at a critical value $\tilde{h}=\tilde{h}_{*}$ as $\tilde{h}$ is increased. 
Conversely, when they are both outside the unit circle at $\tilde{h}=0$, one of the poles moves inside the unit circle at $\tilde{h}_{*}$.
In the former (latter) case, the winding number changes from $+2$ ($-2$) to $0$ at $\tilde{h}_{*}$, leading to the topological transition. 
Figure~\ref{fig:poles} shows the $\tilde{h}$ dependences of $\lvert w_\pm\rvert$ of the exceptional point indicated by the white triangle in Figs.~\ref{fig:topo_trans}(b) and \ref{fig:topo_trans}(d), 
which corresponds to the former case. We derive the concrete formula of $\tilde{h}_{*}$ analytically and discuss the conditions for the topological transition in Appendix~\ref{app:h_star}. 

Thus, our non-Hermitian Hamiltonian in Eq.~\eqref{eq:model_k} exhibits topological transitions with the increase of the magnetic field. 
This provides an interesting example of topological transitions induced by the magnetic field, where the winding number of the exceptional points changes within the gapless phase. 
We will discuss this issue in Sec.~\ref{subsec:discuss_topo}. 

\subsection{Non-Hermitian skin effect}\label{subsec:skin_eff}
\begin{figure}
    \centering
     \includegraphics[width=\columnwidth,clip]{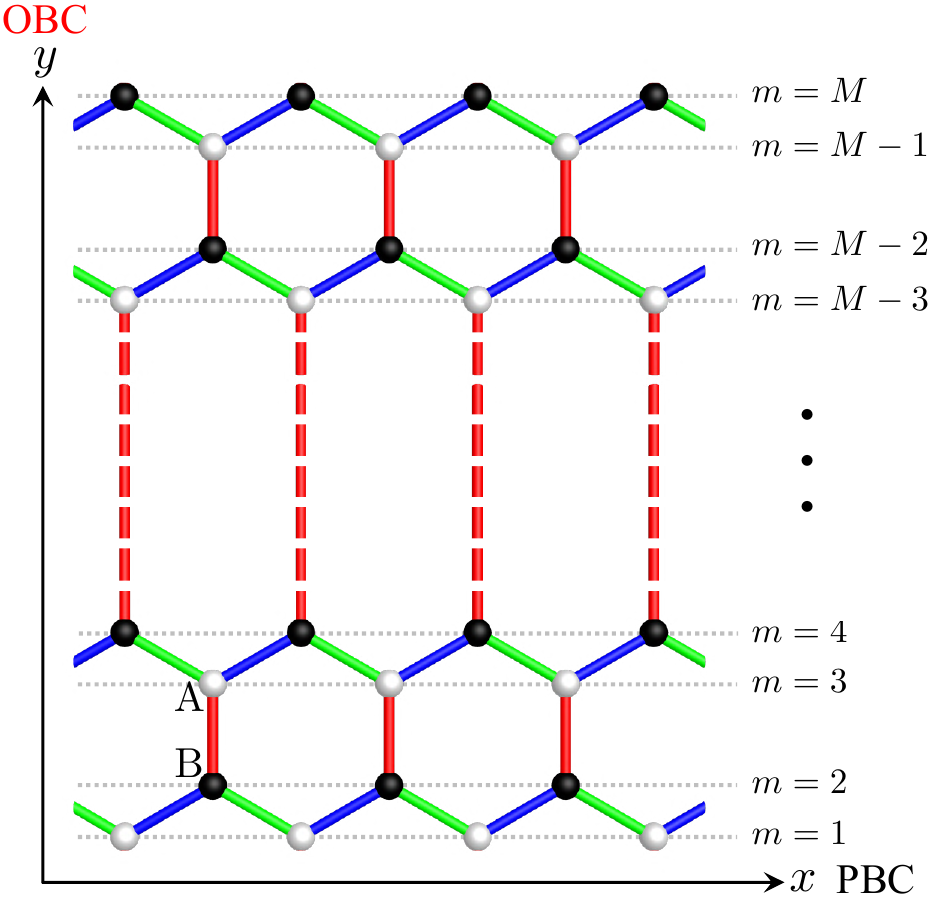}
    \caption{Schematic picture of the system with the open boundary condition (OBC) in the $y$ direction, which leaves zigzag edges. We take the periodic boundary condition (PBC) in the $x$ direction. $M$ denotes the number of the layers.
}
    \label{fig:open}
\end{figure}

\begin{figure*}
    \centering
     \includegraphics[width=\linewidth,clip]{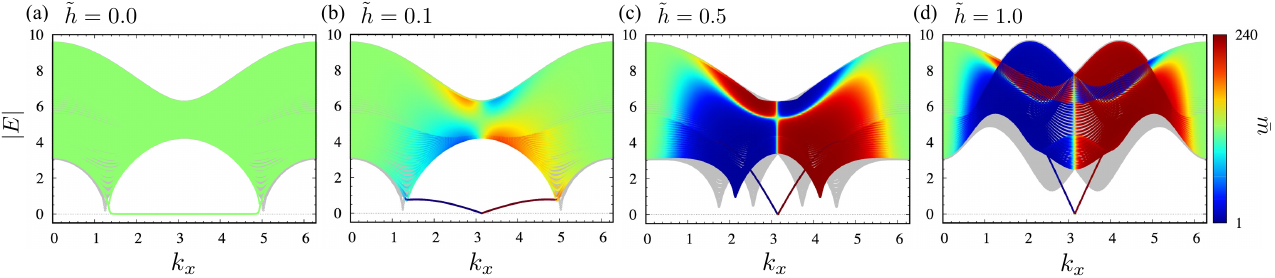}
    \caption{$k_{x}$ dependences of the absolute value of the complex energy eigenvalue, $\lvert E\rvert$, for the system under the PBC in the $x$ direction and the OBC in the $y$ direction at $G_{x}=2.0$, $G_{y}=1.0$, and $G_{z}=2.5\mathrm{e}^{\mathrm{i}\frac{\pi}{3}}$ with (a) $\tilde{h}=0.0$, (b) $\tilde{h}=0.1$, (c) $\tilde{h}=0.5$, and (d) $\tilde{h}=1.0$. We take $M=240$ and $1000$ $k_x$-mesh in the first Brillouin zone.
The color bar shows the center of mass of the wave function in the $y$ direction, $\bar{m}$ in Eq.~\eqref{eq:mtilde}. The data for the system with PBCs in both directions are shown in gray, where we take 1000 $k_{x}$- and 480 $k_{y}$-mesh in the first Brillouin zone.
}
    \label{fig:ed1}
\end{figure*}

\begin{figure*}
    \centering
     \includegraphics[width=\linewidth,clip]{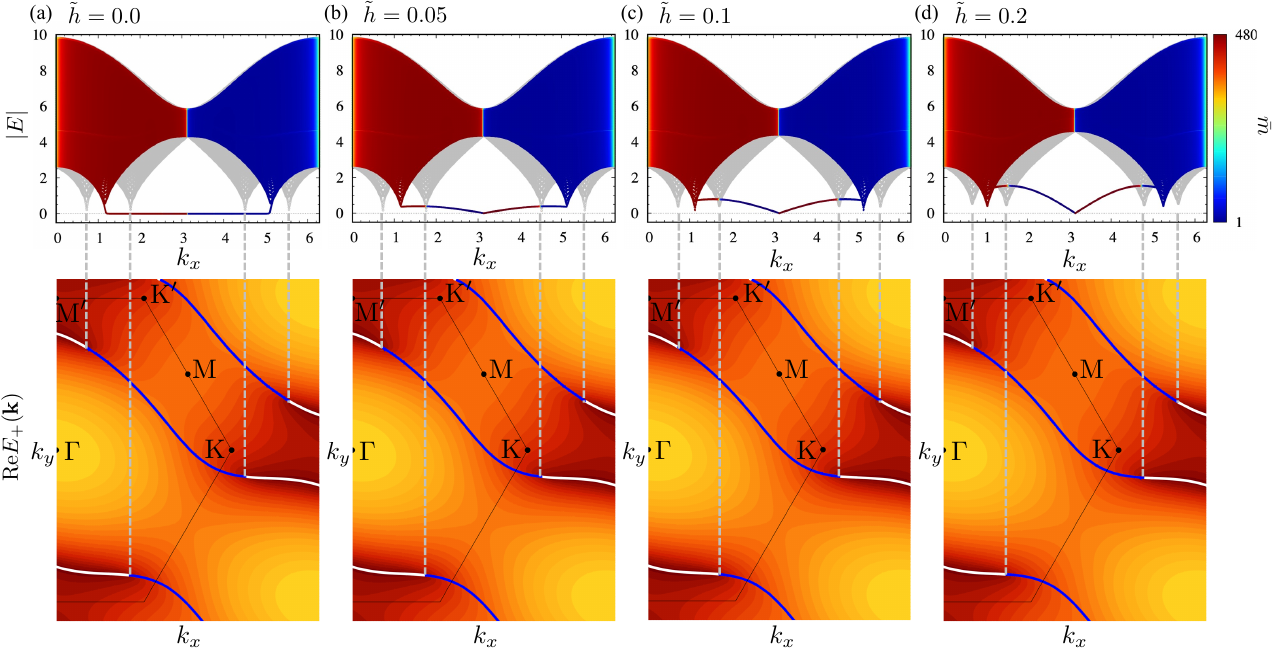}
    \caption{(Upper panels) $k_{x}$ dependences of $\lvert E\rvert$ at $G_{x}=2.0\mathrm{e}^{\mathrm{i}\frac{\pi}{3}}$, $G_{y}=1.0\mathrm{e}^{\mathrm{i}\frac{\pi}{6}}$, and $G_{z}=2.5$, and (a) $\tilde{h}=0.0$, (b) $\tilde{h}=0.05$, (c) $\tilde{h}=0.1$, and (d) $\tilde{h}=0.2$. 
The notations are common to those in Fig.~\ref{fig:ed1}. We take the same $M$ and $\mathbf{k}$-mesh as Fig.~\ref{fig:ed1}. 
Due to numerical precision, we obtain the zero-energy mode in (a) by taking $M=220$ and plot it with $\bar{m}$  
rescaled to $M=480$.
(Lower panels) $\mathrm{Re}E_{+}(\mathbf{k})$ for the system at the same parameters with the PBCs in both directions. The notations are common to those in Fig.~\ref{fig:dispersion1}. The gray dashed lines connect the position of the exceptional points in the lower panels and the zeros in the upper panels. At these values of $k_{x}$, the edge modes switch their chirality, as shown by the color changes between blue and red in the upper panels.
}
    \label{fig:ed2}
\end{figure*}

Next, we study the effect of a magnetic field on the non-Hermitian skin effect by introducing the OBC in one direction and retaining the PBC in the other direction. 
The setup is shown in Fig.~\ref{fig:open}. By cutting the $z$ bonds colored red, we make zigzag open edges at the top and bottom of the system in the $y$ direction. 
Performing the Fourier transformation in the $x$ direction, we can write the Hamiltonian in Eq.~\eqref{eq:model} as
\begin{equation}\label{eq:model_obc}
\mathcal{H}_{\mathrm{eff}}=\sum_{k_x>0}\sum_{m, m'=1}^{M}H_{m', m}(k_{x})c_{-k_{x}, m'}c_{k_{x}, m},
\end{equation}
where
\begin{equation}
H(k_{x})=\begin{pmatrix}
m(k_{x}) & \mathrm{i}r(k_{x}) & d(k_{x}) & 0 & \cdots\\
-\mathrm{i}r(-k_{x}) & -m(k_{x}) & -\mathrm{i}t & -d(k_{x}) & \cdots \\
-d(-k_{x}) & \mathrm{i}t & m(k_{x}) & \mathrm{i}r(k_{x}) & \ddots \\
0 & d(-k_{x}) & -\mathrm{i}r(-k_{x}) & -m(k_{x}) & \ddots\\
\vdots & \vdots & \ddots & \ddots & \ddots
\end{pmatrix}.
\end{equation}
Here, the matrix elements are given by
\begin{align}
r(k_{x}) &= 2(G_{x}\mathrm{e}^{\mathrm{i}k_{x}/2}+G_{y}\mathrm{e}^{-\mathrm{i}k_{x}/2}),\\
t &= 2G_{z},\\
d(k_{x}) &= -4\tilde{h}\sin(k_{x}/2),\\
m(k_{x}) &= 4\tilde{h}\sin(k_{x}).
\end{align}
In the previous study~\cite{Yang2021}, the condition for the non-Hermitian skin effect for the zero magnetic field case was clarified: The skin effect occurs when $\phi_{x}-\phi_{y}\neq0$ (mod $\pi$), where $\phi_{\mu}$ is the complex phase of $G_{\mu}$, in the setting shown in Fig.~\ref{fig:open}. 

Figure~\ref{fig:ed1} shows the absolute value of the complex energy eigenvalues of the system, $\lvert E\rvert$, at $G_{x}=2.0$, $G_{y}=1.0$, and $G_{z}=2.5\mathrm{e}^{\mathrm{i}\frac{\pi}{3}}$, obtained by the numerical diagonalization of Eq.~\eqref{eq:model_obc}. The accuracy of the diagonalization of the non-Hermitian matrices is significantly reduced when the skin effect occurs, and hence, we perform our calculations with quadruple accuracy by using QEISPACK package~\cite{qeispack}. 
Due to the particle-hole symmetry of the system, both states with the energy $E$ and $-E$ are eigenstates, but here we only show the states with $\mathrm{Re}E\geq0$.
The color indicates the average localization of the wave function defined as
\begin{equation}
\bar{m}(n, k_{x})=\sum_{m=1}^{M}m\lvert\psi_{n}(k_{x}, m)\rvert^{2},
\label{eq:mtilde}
\end{equation}
where $\psi_{n}(k_{x}, m)$ is the normalized wave function of the $n$th eigenvalue at ($k_{x}$, $m$). 
For comparison, we also plot $\lvert E_+\rvert$ for the system with PBCs in both directions in gray. 
For this parameter setting with $\phi_x=\phi_y=0$, the skin effect does not occur for $\tilde{h}=0$ as shown in the previous study~\cite{Yang2021}. Indeed, in our result for $\tilde{h}=0$ shown in Fig.~\ref{fig:ed1}(a), no skin effect occurs as all the eigenstates are colored green. 
 In this case, the spectra for the OBC-PBC and PBC-PBC cases coincide with each other, except for the modes lying at $\lvert E\rvert=0$ in the former case. 
We confirm that the zero-energy modes are edge modes by calculating the average square position;
see Appendix~\ref{app:delta_m2}. For $\tilde{h}> 0$, however, the skin effect occurs as shown in Figs.~\ref{fig:ed1}(b)--\ref{fig:ed1}(d), and the difference between the OBC-PBC and PBC-PBC cases becomes pronounced with the increase of $\tilde{h}$. The eigenstates colored blue are localized at the lower edge of the system, while those colored red are localized at the upper edge. We also find that the complex energy spectrum exhibits a change in the point-gap topology corresponding to the induction of the skin effect; see Appendix~\ref{app:comp_spec}.
These indicate that the skin effect is induced by the application of the magnetic field, even for the parameters where it is absent at zero magnetic field.

We note that this behavior appears to be opposite to the suppression of the skin effect by a magnetic field found in the previous study~\cite{Lu2021}. 
This contrasting behavior is presumably due to different nature of the system: The previous study considered models for charged particles such as electrons, incorporating the effect of a magnetic field through the vector potential acting on moving particles, while our present study is for charge-neutral models in which the effect of the magnetic field appears as the sublattice-dependent second-neighbor hoppings of the Majorana fermions.

We also note that the zero-energy edge modes at $\tilde{h}=0$ are colored green, meaning that they are equally localized at both the upper and lower edges of the system (see Appendix~\ref{app:delta_m2}).  
When we turn on $\tilde{h}$, these modes are lifted from zero energy except for $k_{x}=\pi$, and colored blue and red for $k_x<\pi$ and $k_x>\pi$, respectively. This indicates that they are chiral edge modes whose localization direction is different for $k_{x}<\pi$ and $k_{x}>\pi$. These behaviors are similar to those of the edge modes in the Hermitian case. When the dissipation vanishes, these chiral edge modes are continuously connected to those in the Hermitian systems.

Next, we show the results for a parameter where the skin effect occurs for $\tilde{h}=0$ in Fig.~\ref{fig:ed2}. Here we choose $G_{x}=2.0\mathrm{e}^{\mathrm{i}\frac{\pi}{3}}$, $G_{y}=1.0\mathrm{e}^{\mathrm{i}\frac{\pi}{6}}$, and $G_{z}=2.5$, and thus $\phi_{x}-\phi_{y}\neq0$. We can see in Fig.~\ref{fig:ed2}(a) that the skin effect already occurs for $\tilde{h}=0$ as expected, and that the OBC-PBC and PBC-PBC spectra are clearly different. When $\tilde{h}$ is introduced, the bulk spectrum does not change significantly, but the edge modes are modified; not only they are lifted from zero energy except at $k_x=\pi$, but also the localization direction switches at certain $k_{x} \neq \pi$, as shown in Figs.~\ref{fig:ed2}(b)--\ref{fig:ed2}(d). See also Appendices~\ref{app:comp_spec} and \ref{app:delta_m2}. This is regarded as the switching of chirality of the edge modes. Notably, the wavenumbers for this chirality switching correspond to the exceptional points of the PBC-PBC spectra, as shown by the gray dashed lines in Fig.~\ref{fig:ed2}. We will discuss this interesting correspondence in analogy with three-dimensional Hermitian Weyl semimetals in Sec.~\ref{subsec:discuss_Weyl}.

\section{Discussion}\label{sec:discussion}

\subsection{Gapless-gapless topological transitions by a magnetic field}\label{subsec:discuss_topo}
In Sec.~\ref{subsec:topo_trans}, we found interesting topological transitions with changes in the winding number of the exceptional points caused by the magnetic field. 
It is worth noting that the exceptional points whose winding numbers change [indicated by the white triangles in Figs.~\ref{fig:topo_trans}(b) and \ref{fig:topo_trans}(d)] are gapless throughout the topological transition, while new exceptional points are created in pairs [indicated by the white dots in Fig.~\ref{fig:topo_trans}(d)].
This is qualitatively different from the topological transitions studied for other non-Hermitian systems~\cite{Leykam2017, Okugawa2021, Delplace2021, Yoshida2022}. In these cases, while the winding number or a topological number called vorticity~\cite{Leykam2017, Shen2018} changes between two gapless phases, the topological transitions take place only when a gap opens once between them.

It is also worth highlighting that the topological transition in this study is driven by the external magnetic field. This is in contrast to other non-Hermitian systems where the topological transitions are caused by changes in the internal model parameters, such as the tight-binding parameters~\cite{Leykam2017, Okugawa2021, Delplace2021, Yoshida2022}.
This difference stems from the fact that for the Kitaev model the magnetic field effectively generates the second-neighbor hoppings of the Majorana fermions, causing a topological transition through the modulation of the tight-binding parameters. 
Thus, our finding offers an interesting example of non-Hermitian topological transitions driven by an external field.

\subsection{Analogy with the three-dimensional Weyl semimetals}\label{subsec:discuss_Weyl}
\begin{figure}
    \centering
     \includegraphics[width=\columnwidth,clip]{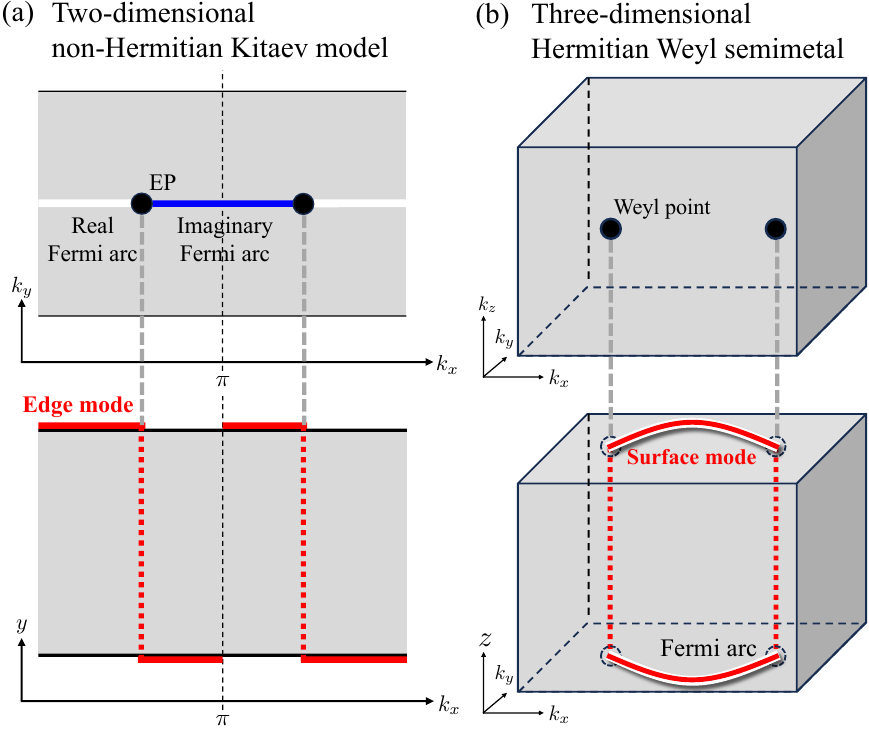}
    \caption{Schematics of (a) edge modes in the two-dimensional non-Hermitian Kitaev model and (b) surface modes in a three-dimensional Hermitian Weyl semimetal, both under a magnetic field. The upper panels of (a) and (b) show the position of EPs and Weyl points in momentum space, respectively. 
    The lower panels of (a) and (b) show switching of the edge and surface modes, respectively. In (a), only one of two chiral edge modes is shown.
    The gray dashed lines in (a) and (b) show the correspondence between the positions of EPs and Weyl points in the upper panels and those of switching points of the edge and surface modes in the lower panels, respectively.
    }
    \label{fig:weyl}
\end{figure}

In Sec.~\ref{subsec:skin_eff}, we found that the chirality of the edge modes is switched at the wavenumbers corresponding to the exceptional points in the bulk spectra. This is similar to the switching of surface Fermi arcs in three-dimensional Hermitian Weyl semimetals under a magnetic field.
In those systems, the surface Fermi arc states switch from one surface to the opposite at wavenumbers corresponding to the projection of the Weyl points in the bulk spectra~\cite{Yan2017, Armitage2018, Zheng2018}. A schematic picture of this similarity is shown in Fig.~\ref{fig:weyl}.
Thus, this similarity provides an example of the analogy between two-dimensional non-Hermitian systems and three-dimensional Hermitian topological systems~\cite{Bergholtz2021}. While this analogy is well known in the PBC cases for general spatial dimensions~\cite{Lee2019e, Schindler2023} due to the number of gap-closing conditions and degree of freedom~\cite{Bergholtz2019, Bergholtz2021}, our result gives a rare example where the analogy holds for the boundary modes in the OBC cases as well.

\subsection{Other contributions within the third-order perturbation}\label{subsec:discuss_perturbation}
In the derivation of the Hamiltonian in Eq.~\eqref{eq:hermite} by the third-order perturbation in terms of the Zeeman coupling, we neglect the second-order contribution and the third-order contribution 
that gives rise to the four-Majorana interaction terms~\cite{Kitaev2006}. Here we discuss the effects of these additional contributions, following the previous study~\cite{Nakazawa2022}.

The second-order contribution is given by
\begin{align}\label{eq:second}
\mathcal{H}_{2}=-\delta\sum_{\langle i, j\rangle_{\mu}}h_{\mu}^{2}\sigma^{\mu}_{i}\sigma^{\mu}_{j}
=\mathrm{i}\delta\sum_{\langle i, j\rangle_{\mu}}h_{\mu}^{2}c_{i}c_{j},
\end{align}
where $\delta$ is assumed to be a constant. This correction gives a bond-dependent change in the Kitaev interaction $J_{\mu}$.
Meanwhile, the third-order contribution that we neglect in Eq.~\eqref{eq:hermite} is given by
\begin{align}\label{eq:third}
\mathcal{H}_{4}
=-\tilde{h}\sum_{[i, j, k]_{\mu,\nu,\lambda}}\sigma^{\mu}_{i}\sigma^{\nu}_{j}\sigma^{\lambda}_{k}
=-\tilde{h}\sum_{[i,j,k; l]}(-1)^{l} c_{i}c_{j}c_{k}c_{l},
\end{align}
where the factor $(-1)^{l}$ takes $+1$ ($-1$) when site $l$ belongs to A (B) sublattice. The summation of $[i, j, k]_{\mu,\nu,\lambda}$ runs over triplets of second-neighbor sites $i$, $j$, and $k$ connected to the common neighboring site by $\mu$, $\nu$, and $\lambda$ bond, respectively ($\mu\neq\nu\neq\lambda\neq\mu$), while the summation of $[i,j,k; l]$ runs over triplets of second-neighbor
sites $i$, $j$, and $k$ connected to the common neighboring site $l$. 
Equation~\eqref{eq:third} includes the four-Majorana interaction terms, which hamper the exact solvability of the problem. Within a mean-field approximation, it leads to modifications of both $J_{\mu}$ and $\tilde{h}$ in Eq.~\eqref{eq:hermite}. Another possible effect of the four-Majorana interactions is an instability of the Majorana Fermi surfaces~\cite{Chari2021}, which appear in the presence of the non-Kitaev interactions~\cite{Takikawa2019}, electric fields~\cite{Chari2021}, or staggard magnetic fields~\cite{Nakazawa2022}.

Summarizing Eqs.~\eqref{eq:second} and \eqref{eq:third} within the mean-field approximation, we obtain the Hamiltonian in the form of Eq.~\eqref{eq:model} with normalized $G_{\mu}$ and $\tilde{h}$ as
\begin{align}
G_{\mu}&\rightarrow G_{\mu}+\delta h_{\mu}^{2}-\tilde{h}(\Delta_{\mathrm{AA}}+\Delta_{\mathrm{BB}}),\\
\tilde{h}&\rightarrow\tilde{h}(1+\Delta_{\mathrm{AB}}),
\end{align}
where we assume spatially uniform mean fields $\Delta_{\mathrm{AA}}=\mathrm{i}\langle c_{i\in \mathrm{A}}c_{j\in\mathrm{A}}\rangle$, $\Delta_{\mathrm{BB}}=\mathrm{i}\langle c_{i\in \mathrm{B}}c_{j\in\mathrm{B}}\rangle$, and $\Delta_{\mathrm{AB}}=\mathrm{i}\langle c_{i\in \mathrm{A}}c_{j\in\mathrm{B}}\rangle$ to decouple the four-Majorana interactions in Eq.~\eqref{eq:third}. 
We note that $\Delta_{\mathrm{AB}}$ is estimated as $\sim-0.52$ in the zero magnetic field~\cite{Baskaran2007, Nasu2015, Knolle2018, Li2018}, and $\Delta_{\mathrm{AA}}=\Delta_{\mathrm{BB}}$ are estimated as between $-0.30$ and $-0.20$ at $\tilde{h}=1.0$~\cite{Li2018, Takahashi2021}, for the Hermitian case with isotropic $J_{x}=J_{y}=J_{z}=1$.
We emphasize that incorporating these corrections does not qualitatively change our results, only shifts the values of $J_{\mu}$ and $\tilde{h}$. 

\section{Summary}\label{sec:summary}
To summarize, we have studied the effect of a magnetic field on the non-Hermitian Kitaev model, which effectively describes the Kitaev model coupled to the environment, based on the perturbation theory with respect to the field strength. We showed that the exceptional points remain gapless up to a nonzero critical field, in stark contrast to the Hermitian case where the Dirac-like nodes are gapped out immediately by applying the magnetic field. We also clarified that the value of the critical effective field depends on the coupling constants, grows beyond the field range where the perturbation theory is valid, and even diverges for some particular parameter conditions on the complex plane of the coupling constants. By calculating the winding number of the exceptional points, we found the topological transition caused by the effective magnetic field within the gapless phase. This offers an interesting example of non-Hermitian topological transitions between two gapless states without opening a gap driven by an external field. In addition, in the system with edges, we showed that the non-Hermitian skin effect can be induced by applying the magnetic field even for the parameters where it does not occur at zero field. We also clarified that the chirality of the edge modes is switched at the wavenumbers corresponding to the exceptional points in the bulk spectrum. This is analogous to the switching of the Fermi arc surface states through the projections of the Weyl points in three-dimensional Weyl semimetals, suggesting a correspondence between the $d$-dimensional non-Hermitian systems and the $(d + 1)$-dimensional Hermitian topological systems. 

Our results provide us with a new possible route to stabilize gapless topological quantum spin liquids under the magnetic field in the presence of dissipation. 
In the particular case, it contains an intriguing example of the topological transitions without gap opening. 
Our results also provide new insight into the connection between the topology of the quantum spin liquids and their edge modes in the presence of dissipation. 
In particular, the finding of the chirality switching of the edge modes suggests the possibility of controlling the direction of the edge modes by dissipation. Thus, our results pave the way for extending the physics of quantum spin liquids to open quantum systems. Further research, especially on other types of quantum spin liquids, is desired.

\begin{acknowledgments}
K.F. thanks K. Kawabata, S. Kitahama, H. Obuse, K. Shimizu, K. Sone, and M. Ueda for constructive suggestions.
The authors thank T. Misawa, J. Nasu, and T. Okubo for fruitful discussions.
Parts of the numerical calculations have been done using the facilities of the Supercomputer Center, the Institute for Solid State Physics, the University of Tokyo, the Information Technology Center, the University of Tokyo, and the Center for Computational Science, University of Tsukuba.
This work was supported by Japan Society for the Promotion of Science (JSPS) KAKENHI Grant Nos. 19H05825, 20H00122, and 22K03509.
\end{acknowledgments}
\appendix
\section{Complex energy spectra} \label{app:comp_spec}
\begin{figure*}
    \centering
     \includegraphics[width=\linewidth,clip]{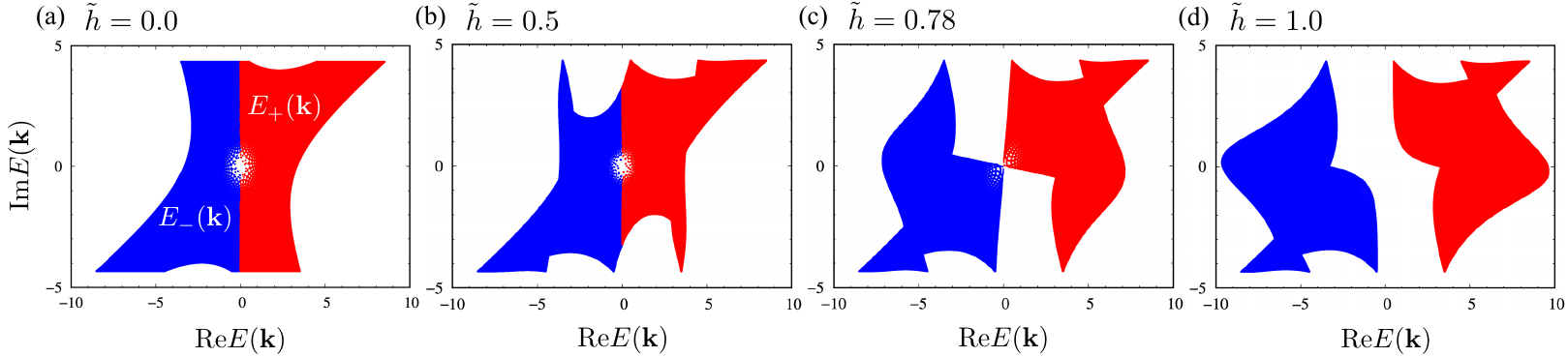}
    \caption{Energy eigenvalues in Eq.~\eqref{eq:eigenval} on the complex plane with $G_{x}=2.0$, $G_{y}=1.0$, and $G_{z}=2.5\mathrm{e}^{\mathrm{i}\frac{\pi}{3}}$ under the effective magnetic field of (a) $\tilde{h}=0.0$, (b) $\tilde{h}=0.5$, (c) $\tilde{h}=0.78$, and (d) $\tilde{h}=1.0$. The red and blue points belong to the energy eigenvalues in $E_{+}(\mathbf{k})$ and $E_{-}(\mathbf{k})$, respectively. We use $500\times1000$ mesh in the half of the first Brillouin zone.
}
    \label{fig:pbc_spec}
\end{figure*}

\begin{figure*}
    \centering
     \includegraphics[width=\linewidth,clip]{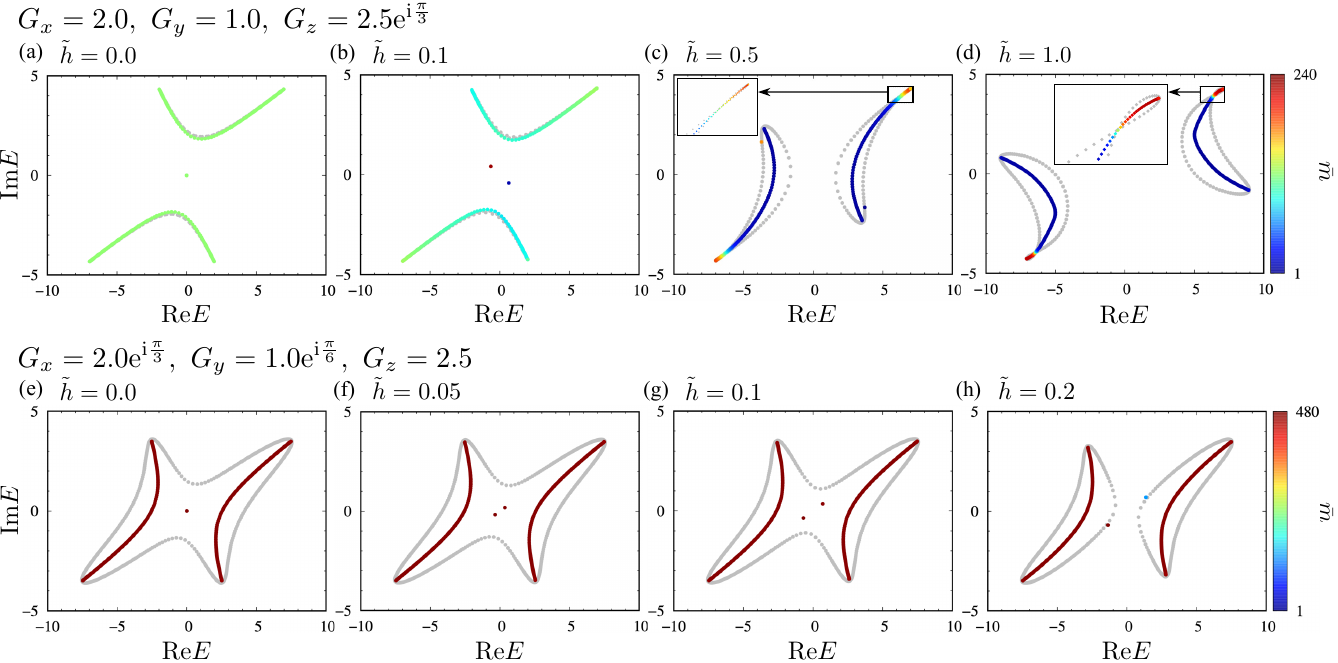}
    \caption{Energy eigenvalues of the Hamiltonian in Eq.~\eqref{eq:model_obc} on the complex plane at $G_{x}=2.0$, $G_{y}=1.0$, $G_{z}=2.5\mathrm{e}^{\mathrm{i}\frac{\pi}{3}}$, and (a) $\tilde{h}=0.0$, (b) $\tilde{h}=0.1$, (c) $\tilde{h}=0.5$, (d) $\tilde{h}=1.0$, and at $G_{x}=2.0\mathrm{e}^{\mathrm{i}\frac{\pi}{3}}$, $G_{y}=1.0\mathrm{e}^{\mathrm{i}\frac{\pi}{6}}$, $G_{z}=2.5$, and (e) $\tilde{h}=0.0$, (f) $\tilde{h}=0.05$, (g) $\tilde{h}=0.1$, and (h) $\tilde{h}=0.2$. Data with the OBC in the $y$ direction are shown in color indicating the center of mass of the wave function $\bar{m}$ in Eq.~\eqref{eq:mtilde}, while those with the PBC are shown in gray. For both data, we adopt the PBC in the $x$ direction and fix $k_{x}=\pi/2$. We use the same system size as Figs.~\ref{fig:ed1} and \ref{fig:ed2} for (a)--(d) and (e)--(h), respectively. Inset of (c) and (d) show enlarged view of the region indicated by the black rectangle.
}
    \label{fig:obc_spec}
\end{figure*}

In this Appendix, we show supplemental data on the complex energy spectra corresponding to Figs.~\ref{fig:dispersion1}, \ref{fig:ed1}, and \ref{fig:ed2}. Figure~\ref{fig:pbc_spec} shows the energy eigenvalues in Eq.~\eqref{eq:eigenval} on the complex plane. Figs.~\ref{fig:pbc_spec}(a), \ref{fig:pbc_spec}(b), and \ref{fig:pbc_spec}(d) correspond to Figs.~\ref{fig:dispersion1}(a), \ref{fig:dispersion1}(b), and \ref{fig:dispersion1}(c), respectively. A clear difference in the complex energy spectrum can be seen between the gapless [Figs.~\ref{fig:pbc_spec}(a) and \ref{fig:pbc_spec}(b)] and gapped states [Fig.~\ref{fig:pbc_spec}(d)]. At $\tilde{h}=0.78$ in Fig.~\ref{fig:pbc_spec}, which is very close to the critical effective magnetic field $\tilde{h}_{\mathrm{c}}$, we can see the gap opening behavior. Note that the white dots around $E(\mathbf{k})=0$ in Figs.~\ref{fig:pbc_spec}(a), \ref{fig:pbc_spec}(b), and \ref{fig:pbc_spec}(c) are due to  the finite size effects.

Figures~\ref{fig:obc_spec}(a)--\ref{fig:obc_spec}(d) show the complex energy eigenvalues corresponding to Fig.~\ref{fig:ed1} with $k_{x}=\pi/2$. Note that we show the eigenvalues with both $\mathrm{Re}E\geq 0$ and $\mathrm{Re}E<0$, although we only show those with $\mathrm{Re}E\geq 0$ in Fig.~\ref{fig:ed1} and the upper panels of Fig.~\ref{fig:ed2}.
At $\tilde{h}=0$ in Fig.~\ref{fig:obc_spec}(a), all OBC eigenvalues are colored green, which means that the wave functions of all eigenmodes are spread over the entire system, except for the edge modes indicated by the isolated point at $E=0$ (doubly degenerate; see also Appendix~\ref{app:delta_m2}. They turn red or blue as $\tilde{h}$ is increased in Figs.~\ref{fig:obc_spec}(b)--\ref{fig:obc_spec}(d), suggesting the wave functions are localized at only the top or bottom edge, respectively, as found in Sec.~\ref{subsec:skin_eff}. 
The induction of the non-Hermitian skin effect can be understood as a change in the point-gap topology of the complex spectra~\cite{Kawabata2019, Ding2022, Okuma2022, Zhang2022a, Lin2023}. Corresponding to the induction of the skin effect in Figs.~\ref{fig:ed1}(c) and \ref{fig:ed1}(d), the PBC spectra enclose finite areas in Figs.~\ref{fig:obc_spec}(c) and \ref{fig:obc_spec}(d). In addition, the energy dependency of localization direction can be understood as the twist of the PBC spectrum. The points where the localization direction changes in the OBC spectrum correspond to the points where the PBC spectrum intersects itself on the complex plane, as shown in Figs.~\ref{fig:obc_spec}(c) and \ref{fig:obc_spec}(d).
Figures~\ref{fig:obc_spec}(e)--\ref{fig:obc_spec}(h) show the results corresponding to Fig.~\ref{fig:ed2} with $k_{x}=\pi/2$. All OBC eigenvalues except for the edge modes are localized at the upper edge of the system. In Figs.~\ref{fig:obc_spec}(e)--\ref{fig:obc_spec}(g), edge modes are also localized at the upper edge, however, one of the two edge modes is localized at the bottom edge in Fig.~\ref{fig:obc_spec}(h).
 It reflects the switching of the edge modes discussed in Sec.~\ref{subsec:skin_eff}.

\section{Conditions for divergence of $\tilde{h}_{\mathrm{c}}$}\label{app:ring}
In this Appendix, we derive the divergence conditions of $\tilde{h}_{\mathrm{c}}$ in Eqs.~\eqref{eq:ring_xyz},
\eqref{eq:dots_xyz}, and \eqref{eq:dot_xyz}. 
The conditions are obtained by noting that the energy dispersion in Eq.~\eqref{eq:eigenval} remains gapless at the exceptional points satisfying $\varepsilon(\mathbf{k})=0$ or $\varepsilon(-\mathbf{k})=0$, if $\Delta(\mathbf{k})=0$ for $\tilde{h}>0$. There are three possibilities to fulfill this situation: 
(i) $\tilde{k}_{1}=\tilde{k}_{2}$ (mod $2\pi$),
(ii) $\tilde{k}_{1}\neq0$ and $\tilde{k}_{2}=0$ (mod $2\pi$),
and (iii) $\tilde{k}_{1}=0$ and $\tilde{k}_{2}\neq0$ (mod $2\pi$),
where $\tilde{k}_{1}=\mathbf{k}\cdot\mathbf{a}_{1}$ and $\tilde{k}_{2}=\mathbf{k}\cdot\mathbf{a}_{2}$. 

For the case (i), from $\varepsilon(\mathbf{k})=0$ in Eq.~\eqref{eq:epsilon_k}, i.e., $\mathrm{Re}\varepsilon(\mathbf{k})=0$ and $\mathrm{Im}\varepsilon(\mathbf{k})=0$, we obtain
\begin{equation}
\begin{pmatrix} c & -s \\ s & c \end{pmatrix}
\begin{pmatrix} \cos\tilde{k}_{1} \\ \sin\tilde{k}_{1} \end{pmatrix}
+\begin{pmatrix} u \\ v \end{pmatrix}
=0,
\end{equation}
where $c=\lvert G_{x}\rvert\cos\phi_{x}+\lvert G_{y}\rvert\cos\phi_{y}$, $s=\lvert G_{x}\rvert\sin\phi_{x}+\lvert G_{y}\rvert\sin\phi_{y}$, and $G_{z}=u+\mathrm{i}v$. 
This leads to 
\begin{equation}\label{eq:condition_1}
\cos\tilde{k}_{1}=-\frac{cu+sv}{c^{2}+s^{2}},\ \sin\tilde{k}_{1}=\frac{su-cv}{c^{2}+s^{2}}.
\end{equation}
Combing these relations with $\cos^{2}\tilde{k}_{1}+\sin^{2}\tilde{k}_{1}=1$, we obtain the condition 
\begin{equation}\label{eq:ring_x_uv}
u^{2}+v^{2}=c^{2}+s^{2}, 
\end{equation}
which corresponds to the ring given by $\lvert G_{z}\rvert=\lvert G_{x}+G_{y}\rvert$ in Eq.~\eqref{eq:ring_xyz}. 
Similarly, for $\varepsilon(-\mathbf{k})=0$, we obtain
\begin{equation}\label{eq:condition_2}
\cos\tilde{k}_{1}=-\frac{cu+sv}{c^{2}+s^{2}},\ \sin\tilde{k}_{1}=-\frac{su-cv}{c^{2}+s^{2}},
\end{equation}
which leads to the same condition. 
When $\varepsilon(\mathbf{k})=0$ and $\varepsilon(-\mathbf{k})=0$ are satisfied simultaneously, 
Eqs.~\eqref{eq:condition_1} and \eqref{eq:condition_2} give $su-cv=0$. This condition and Eq.~\eqref{eq:ring_x_uv} leads to 
\begin{equation}\label{eq:uv_dot_xyz}
(u, v)=(\pm c, \pm s).
\end{equation}
The upper and lower signs correspond to the last equation in Eq.~\eqref{eq:dots_xyz} and Eq.~\eqref{eq:dot_xyz}, respectively.

For the case (ii), we can rewrite Eq.~\eqref{eq:epsilon_k} as $\varepsilon(\mathbf{k})=2\mathrm{e}^{\mathrm{i}\phi_{y}}\varepsilon'$ with
\begin{align}
\varepsilon' =\lvert G_{x}\rvert\mathrm{e}^{\mathrm{i}\tilde{k}_{1}+\mathrm{i}\Delta\phi}+\lvert G_{y}\rvert + u'+\mathrm{i}v',
\end{align}
where $\Delta\phi=\phi_{x}-\phi_{y}$ and $\lvert G_{z}\rvert\mathrm{e}^{\mathrm{i}\phi_{z}-\mathrm{i}\phi_{y}}=u'+\mathrm{i}v'$. To satisfy $\varepsilon(\mathbf{k})=0$, we need $\mathrm{Re}\varepsilon'=0$ and $\mathrm{Im}\varepsilon'=0$, which result in 
\begin{align}\label{eq:condition_3}
\cos(\tilde{k}_{1}+\Delta\phi)=-\frac{\lvert G_{y}\rvert +u'}{\lvert G_{x}\rvert},
\end{align}
and
\begin{align}\label{eq:condition_4}
\sin(\tilde{k}_{1}+\Delta\phi)=-\frac{v'}{\lvert G_{x}\rvert},
\end{align}
respectively.
From these two and $\cos^{2}(\tilde{k}_{1}+\Delta\phi)+\sin^{2}(\tilde{k}_{1}+\Delta\phi)=1$, we obtain the condition 
\begin{equation}
(u'+\lvert G_{y}\rvert)^{2}+v'^{2}=\lvert G_{x}\rvert^{2},
\label{eq:condition_8}
\end{equation}
which corresponds to the ring given by $\lvert G_{x}\rvert=\lvert G_{y}+G_{z}\rvert$ in Eq.~\eqref{eq:ring_xyz}. 
For the case with $\varepsilon(-\mathbf{k})=0$, we obtain similar relations 
\begin{align}
\cos(-\tilde{k}_{1}+\Delta\phi)&=-\frac{\lvert G_{y}\rvert +u'}{\lvert G_{x}\rvert}, \label{eq:condition_5}\\
\sin(-\tilde{k}_{1}+\Delta\phi)&=-\frac{v'}{\lvert G_{x}\rvert}, \label{eq:condition_6}
\end{align}
which leads to the same condition as Eq.~\eqref{eq:condition_8}.
When $\varepsilon(\mathbf{k})=0$ and $\varepsilon(-\mathbf{k})=0$ are both satisfied, Eqs.~\eqref{eq:condition_3}, \eqref{eq:condition_4}, \eqref{eq:condition_5}, and \eqref{eq:condition_6} give 
\begin{equation}
(u', v')=(-\lvert G_{y}\rvert\pm\lvert G_{x}\rvert\cos\Delta\phi, \pm\lvert G_{x}\rvert\sin\Delta\phi).
\end{equation}
The upper and lower signs give the first equation in Eq.~\eqref{eq:dots_xyz} and Eq.~\eqref{eq:dot_xyz}, respectively.

Finally, for the case (iii), by replacing $\tilde{k}_{1}$ by $\tilde{k}_{2}$ and interchanging $G_{x}$ and $G_{y}$ in the derivation for the case (ii), we obtain 
\begin{equation}\label{eq:condition_7}
(u'+\lvert G_{x}\rvert)^{2}+v'^{2}=\lvert G_{y}\rvert^{2},
\end{equation}
which corresponds to the ring given by $\lvert G_{y}\rvert=\lvert G_{z}+G_{x}\rvert$ in Eq.~\eqref{eq:ring_xyz}.  
In addition, we can also obtain the condition for the point satisfying $\varepsilon(\mathbf{k})=\varepsilon(-\mathbf{k})=0$:
\begin{equation}
(u', v')=(-\lvert G_{x}\rvert\pm\lvert G_{y}\rvert\cos\Delta\phi, \mp\lvert G_{y}\rvert\sin\Delta\phi).
\end{equation}
The upper and lower signs correspond to the second equation in Eq.~\eqref{eq:dots_xyz} and Eq.~\eqref{eq:dot_xyz}, respectively.

\section{Calculations of the winding number} \label{app:winding}
In this Appendix, we calculate the winding number of the exceptional points in Eq.~\eqref{eq:winding} analytically. From the explicit form of the Majorana Bloch Hamiltonian in Eq.~\eqref{eq:model_k}, the integrand can be expressed as
\begin{align}
&\nabla_{\mathbf{k}}\mathrm{ln}\det [H(\mathbf{k})] \notag\\
&=\frac{[\nabla_{\mathbf{k}}\varepsilon(\mathbf{k})]\varepsilon(-\mathbf{k})+\varepsilon(\mathbf{k})\nabla_{\mathbf{k}}\varepsilon(-\mathbf{k})+2\Delta(\mathbf{k})\nabla_{\mathbf{k}}\Delta(\mathbf{k})}{\varepsilon(\mathbf{k})\varepsilon(-\mathbf{k})+\Delta(\mathbf{k})^{2}}. \label{eq:integrand}
\end{align}
Note that $E(\mathbf{k}_{\mathrm{EP}})=0$. 
We expand Eq.~\eqref{eq:integrand} near the exceptional point at $\mathbf{k}=\mathbf{k}_{\mathrm{EP}}$ as
\begin{align}
\varepsilon(\mathbf{k})&\simeq \varepsilon(\mathbf{k}_{\mathrm{EP}})+\mathrm{i}A_{+}(k_{x}-k_{\mathrm{EP}, x})+\mathrm{i}B_{+}(k_{y}-k_{\mathrm{EP}, y}), \label{eq:expand_e+}\\
\varepsilon(-\mathbf{k})&\simeq \varepsilon(-\mathbf{k}_{\mathrm{EP}})+\mathrm{i}A_{-}(k_{x}-k_{\mathrm{EP}, x})+\mathrm{i}B_{-}(k_{y}-k_{\mathrm{EP}, y}), \label{eq:expand_e-}\\
\Delta(\mathbf{k})&\simeq\Delta(\mathbf{k}_{\mathrm{EP}})+D_{1}(k_{x}-k_{\mathrm{EP}, x})+D_{2}(k_{y}-k_{\mathrm{EP}, y}), \label{eq:expand_D}
\end{align}
where 
\begin{align}
\label{eq:Apm}
A_{\pm}&=\lvert G_{x}\rvert\mathrm{e}^{\mathrm{i}(\pm\mathbf{k}_{\mathrm{EP}}\cdot\mathbf{a}_{1}+\phi_{x})}-\lvert G_{y}\rvert\mathrm{e}^{\mathrm{i}(\pm\mathbf{k}_{\mathrm{EP}}\cdot\mathbf{a}_{2}+\phi_{y})},\\
B_{\pm}&=\sqrt{3}[\lvert G_{x}\rvert\mathrm{e}^{\mathrm{i}(\pm\mathbf{k}_{\mathrm{EP}}\cdot\mathbf{a}_{1}+\phi_{x}}+\lvert G_{y}\rvert\mathrm{e}^{\mathrm{i}(\pm\mathbf{k}_{\mathrm{EP}}\cdot\mathbf{a}_{2}+\phi_{y})}],\\
D_{1}&=2\tilde{h}[-\cos(\mathbf{k}_{\mathrm{EP}}\cdot\mathbf{a}_{1})-\cos(\mathbf{k}_{\mathrm{EP}}\cdot\mathbf{a}_{2})+2\cos(k_{\mathrm{EP}, x})],\\
D_{2}&=2\sqrt{3}\tilde{h}[-\cos(\mathbf{k}_{\mathrm{EP}}\cdot\mathbf{a}_{1})+\cos(\mathbf{k}_{\mathrm{EP}}\cdot\mathbf{a}_{2})].
\label{eq:D2}
\end{align}

First, we calculate the winding number of the exceptional points at a generic parameter in the B phase except for the rings described by Eq.~\eqref{eq:ring_xyz}. 
Substituting Eqs.~\eqref{eq:expand_e+}--\eqref{eq:expand_D} into Eq.~\eqref{eq:integrand} and neglecting the regular part, Eq.~\eqref{eq:winding} can be written as
\begin{equation}\label{eq:winding_AB}
W=\oint_{C}\frac{1}{2\pi}\frac{A'\mathrm{d}k_{x}+B'\mathrm{d}k_{y}}{A'(k_{x}-k_{\mathrm{EP}, x})+B'(k_{y}-k_{\mathrm{EP}, y})},
\end{equation}
where
\begin{align}
A'&=\mathrm{i}[\varepsilon(-\mathbf{k}_{\mathrm{EP}})A_{+}-\varepsilon(\mathbf{k}_{\mathrm{EP}})A_{-}]+2\Delta(\mathbf{k}_{\mathrm{EP}})D_{1},\\
B'&=\mathrm{i}[\varepsilon(-\mathbf{k}_{\mathrm{EP}})B_{+}-\varepsilon(\mathbf{k}_{\mathrm{EP}})B_{-}]+2\Delta(\mathbf{k}_{\mathrm{EP}})D_{2}.
\end{align}
By parametrizing the integration path $C$ with a circle of radius $r$ around $\mathbf{k}_{\mathrm{EP}}$ as $\mathbf{k}=\mathbf{k}_{\mathrm{EP}}+r(\cos\theta, \sin\theta)$ $(0\leq\theta<2\pi)$, we can rewrite the winding number as 
\begin{align}
W&=\int_{0}^{2\pi}\frac{\mathrm{d}\theta}{2\pi}\frac{-A'\sin\theta+B'\cos\theta}{A'\cos\theta+B'\sin\theta}
\nonumber \\
&=\oint_{\lvert z\rvert=1}\frac{\mathrm{d}z}{2\pi\mathrm{i}}\frac{1}{\mathrm{i}z}\frac{-A'(z^{2}-1)+\mathrm{i}B'(z^{2}+1)}{\mathrm{i}A'(z^{2}+1)+B'(z^{2}-1)} 
\nonumber\\
&=\oint_{\lvert z\rvert=1}\frac{\mathrm{d}z}{2\pi\mathrm{i}}\frac{z^{2}-\frac{A'+\mathrm{i}B'}{A'-\mathrm{i}B'}}{z\left(z-\sqrt{\frac{B'-\mathrm{i}A'}{B'+\mathrm{i}A'}}\right)\left(z+\sqrt{\frac{B'-\mathrm{i}A'}{B'+\mathrm{i}A'}}\right)}. \label{eq:winding_theta}
\end{align}
At the second equality, we change the integral variable from $\theta$ to $z=\mathrm{e}^{\mathrm{i}\theta}$. The integrand in Eq.~\eqref{eq:winding_theta} has three poles at $z=0$, $+\sqrt{\frac{B'-\mathrm{i}A'}{B+\mathrm{i}A'}}$, and $-\sqrt{\frac{B'-\mathrm{i}A'}{B+\mathrm{i}A'}}$ with residues $-1$, $+1$, and $+1$, respectively. Therefore, we obtain
\begin{equation}
W=\begin{cases}
+1 & \left(\left\lvert\sqrt{\frac{B'-\mathrm{i}A'}{B'+\mathrm{i}A'}}\right\rvert<1\right)
\\ 
-1 & \left(\left\lvert\sqrt{\frac{B'-\mathrm{i}A'}{B'+\mathrm{i}A'}}\right\rvert>1 \right).
\end{cases}
\end{equation}

Next, we consider a parameter on the rings in Eq.~\eqref{eq:ring_xyz} except for the points where $\varepsilon(\mathbf{k}_{\mathrm{EP}})=0$ 
and
$\varepsilon(-\mathbf{k}_{\mathrm{EP}})=0$
are satisfied simultaneously.
When $\varepsilon(\mathbf{k}_{\mathrm{EP}})=0$ and $\varepsilon(-\mathbf{k}_{\mathrm{EP}})\neq0$, Eq.~\eqref{eq:winding} can be written as
\begin{equation}
W=\oint_{C}\frac{1}{2\pi}\frac{A_{+}\mathrm{d}k_{x}+B_{+}\mathrm{d}k_{y}}{A_{+}(k_{x}-k_{\mathrm{EP}, x})+B_{+}(k_{y}-k_{\mathrm{EP}, y})}.
\label{eq:W2}
\end{equation}
Note that $\Delta(\mathbf{k}_{\mathrm{EP}})=0$ on the rings described by Eq.~\eqref{eq:ring_xyz}. 
Equation~\eqref{eq:W2} is equivalent to Eq.~\eqref{eq:winding_AB} by replacing $A'$ by $A_{+}$ and $B'$ by $B_{+}$. 
Hence, similar calculations to Eq.~\eqref{eq:winding_theta} yield
\begin{equation}
W=\begin{cases}
+1 & \left(\left\lvert\sqrt{\frac{B_{+}-\mathrm{i}A_{+}}{B_{+}+\mathrm{i}A_{+}}}\right\rvert<1 \right)
\\ 
-1 & \left( \left\lvert\sqrt{\frac{B_{+}-\mathrm{i}A_{+}}{B_{+}+\mathrm{i}A_{+}}}\right\rvert>1 \right). 
\end{cases}
\end{equation}
Similarly, when $\varepsilon(\mathbf{k}_{\mathrm{EP}})\neq0$ and $\varepsilon(-\mathbf{k}_{\mathrm{EP}})=0$, by replacing $A'$ by $A_{-}$ and $B'$ by $B_{-}$ in Eq.~\eqref{eq:winding_AB}, we obtain
\begin{equation}
W=\begin{cases}
+1 & \left( \left\lvert\sqrt{\frac{B_{-}-\mathrm{i}A_{-}}{B_{-}+\mathrm{i}A_{-}}}\right\rvert<1 \right)
\\ 
-1 & \left( \left\lvert\sqrt{\frac{B_{-}-\mathrm{i}A_{-}}{B_{+}+\mathrm{i}A_{-}}}\right\rvert>1 \right). 
\end{cases}
\end{equation}

Finally, we calculate the winding number of the exceptional points described by Eqs.~\eqref{eq:dots_xyz} and \eqref{eq:dot_xyz}, where $\varepsilon(\mathbf{k}_{\mathrm{EP}})=\varepsilon(-\mathbf{k}_{\mathrm{EP}})=0$. 
In this case, using the integral variable $z=\mathrm{e}^{2\mathrm{i}\theta}$, we obtain 
\begin{align}
 W =\oint_{\lvert z\rvert=1}\frac{\mathrm{d}z}{2\pi\mathrm{i}}
 \frac{2}{z}\frac{z^{2}+\frac{F_{1}+\mathrm{i}F_{2}}{F_{1}-\mathrm{i}F_{2}}}{z^{2}+2\mathrm{i}\frac{F_{3}+F_{4}}{F_{1}-\mathrm{i}F_{2}}z-\frac{F_{1}+\mathrm{i}F_{2}}{F_{1}-\mathrm{i}F_{2}}},
\end{align}
where
\begin{align}
F_{1}&=A_{+}B_{-}+A_{-}B_{+}+2D_{1}D_{2},\\
F_{2}&=-A_{+}A_{-}+B_{+}B_{-}-D_{1}^{2}+D_{2}^{2},\\
F_{3}&=A_{+}A_{-}+D_{1}^{2},\\
F_{4}&=B_{+}B_{-}+D_{2}^{2}.
\end{align}
This corresponds to Eq.~\eqref{eq:winding_wpm} with 
\begin{equation}\label{eq:poles}
w_{\pm}=-\mathrm{i}\frac{F_{3}+F_{4}}{F_{1}-\mathrm{i}F_{2}}\pm\sqrt{-\left(\frac{F_{3}+F_{4}}{F_{1}-\mathrm{i}F_{2}}\right)^{2}+\frac{F_{1}+\mathrm{i}F_{2}}{F_{1}-\mathrm{i}F_{2}}}.
\end{equation} 
The poles $z=0$, $w_+$, and $w_-$ have residues $-2$, $+2$, and $+2$, respectively. Hence we conclude
\begin{equation}
W=\begin{cases}
+2 & \left( \lvert w_{+} \rvert,\ \lvert w_{-} \rvert<1 \right)
\\ 
-2 & \left(  \lvert w_{+} \rvert,\ \lvert w_{-} \rvert>1\right)
\\ 
0 & \text{otherwise},
\end{cases}
\end{equation}
as discussed in Sec.~\ref{subsec:topo_trans}. 

\section{Derivation of $\tilde{h}_{*}$} \label{app:h_star}
In this Appendix, we derive the analytic formulas for $\tilde{h}_{*}$ where the topological transition occurs with changes in the winding number. 
The topological transition occurs at the exceptional points satisfying $\varepsilon(\mathbf{k}_{\mathrm{EP}})=0$ and $\varepsilon(-\mathbf{k}_{\mathrm{EP}})=0$ simultaneously, which appear at the parameters described by Eqs.~\eqref{eq:dots_xyz} and \eqref{eq:dot_xyz}.
We begin with the first equation in Eq.~\eqref{eq:dots_xyz}.  
As discussed in Appendix~\ref{app:ring}, the exceptional points satisfying $\varepsilon(\mathbf{k}_{\mathrm{EP}})=\varepsilon(-\mathbf{k}_{\mathrm{EP}})=0$ fulfill the conditions $\tilde{k}_{1}\neq 0$ and $\tilde{k}_{2}=0$ (mod $2\pi$).
From this and the conditions in Eqs.~\eqref{eq:condition_4} and \eqref{eq:condition_6}, we obtain
$\sin(\tilde{k}_{1}+\Delta\phi)=\sin(-\tilde{k}_{1}+\Delta\phi)$, which is satisfied only by $\tilde{k}_{1}=\pi$.
By substituting $\tilde{k}_1=\pi$ and $\tilde{k}_2=0$
to Eqs.~\eqref{eq:Apm}--\eqref{eq:D2}, we obtain
\begin{align}
\label{eq:A+}
A_{+}&=A_{-}=-(\lvert G_{x}\rvert \mathrm{e}^{\mathrm{i}\phi_{x}}+\lvert G_{y}\rvert \mathrm{e}^{\mathrm{i}\phi_{y}}),\\
B_{+}&=B_{-}=-\sqrt{3}(\lvert G_{x}\rvert \mathrm{e}^{\mathrm{i}\phi_{x}}-\lvert G_{y}\rvert \mathrm{e}^{\mathrm{i}\phi_{y}}),\\
D_{1}&=-4\tilde{h},\\
\label{eq:D2}
D_{2}&=4\sqrt{3}\tilde{h}.
\end{align}
The topological transition occurs when $\lvert w_{+}\rvert$ or $\lvert w_{-}\rvert$ becomes unity, as discussed in Appendix~\ref{app:winding}. 
From Eq.~\eqref{eq:poles} with Eqs.~\eqref{eq:A+}--\eqref{eq:D2}, we can derive the explicit formula of $\tilde{h}_{*}$ as
\begin{equation}
\tilde{h}_{*}=\mp\frac{\lvert G_{y}\rvert}{4}\frac{\sin(\phi_{x}-\phi_{y})}{\cos\phi_{x}}.
\label{eq:tildeh1}
\end{equation}

Similarly, we can obtain the fomulas for $\tilde{h}_*$ for the second and third equations in Eq.~\eqref{eq:dots_xyz}. 
For the second equation, Eqs.~\eqref{eq:condition_4} and \eqref{eq:condition_6} with replacing $\tilde{k}_{1}$ by $\tilde{k}_{2}$ and interchanging $G_{x}$ and $G_{y}$ lead to $\tilde{k}_{1}=0$ and $\tilde{k}_{2}=\pi$. 
By substituting them, we obtain
\begin{align}
A_{+}&=A_{-}=\lvert G_{x}\rvert \mathrm{e}^{\mathrm{i}\phi_{x}}+\lvert G_{y}\rvert \mathrm{e}^{\mathrm{i}\phi_{y}},\\
B_{+}&=B_{-}=\sqrt{3}(\lvert G_{x}\rvert \mathrm{e}^{\mathrm{i}\phi_{x}}-\lvert G_{y}\rvert \mathrm{e}^{\mathrm{i}\phi_{y}}),\\
D_{1}&=-4\tilde{h},\\
D_{2}&=-4\sqrt{3}\tilde{h},
\end{align}
and also derive 
\begin{equation}
\tilde{h}_{*}=\mp\frac{\lvert G_{x} \rvert}{4}\frac{\sin(\phi_{x}-\phi_{y})}{\cos\phi_{y}}.
\end{equation}
Meanwhile, for the third equation in Eq.~\eqref{eq:dots_xyz}, we obtain $\tilde{k}_{1}=\tilde{k}_{2}=\pi$ by substituting the upper sign of Eq.~\eqref{eq:uv_dot_xyz} into Eqs.~\eqref{eq:condition_1} and \eqref{eq:condition_2}, and hence, 
\begin{align}
A_{+}&=A_{-}=-(\lvert G_{x}\rvert \mathrm{e}^{\mathrm{i}\phi_{x}}-\lvert G_{y}\rvert \mathrm{e}^{\mathrm{i}\phi_{y}}),\\
B_{+}&=B_{-}=-\sqrt{3}(\lvert G_{x}\rvert \mathrm{e}^{\mathrm{i}\phi_{x}}+\lvert G_{y}\rvert \mathrm{e}^{\mathrm{i}\phi_{y}}),\\
D_{1}&=8\tilde{h},\\
D_{2}&=0,
\end{align}
and
\begin{equation}
\tilde{h}_{*}=\mp\frac{\lvert G_{x}\rvert\lvert G_{y}\rvert\sin(\phi_{x}-\phi_{y})}{4(\lvert G_{x}\rvert\cos\phi_{x}+\lvert G_{y}\rvert\cos\phi_{y})}.
\end{equation}

Finally, we consider the case of the intersection of three rings described by Eq.~\eqref{eq:dot_xyz}. 
By substituting the lower sign of Eq.~\eqref{eq:uv_dot_xyz} into Eqs.~\eqref{eq:condition_1} and \eqref{eq:condition_2}, 
we can obtain the location of the exceptional points as $\tilde{k}_{1}=\tilde{k}_{2}=0$. 
Hence, we obtain
\begin{align}
A_{+}&=A_{-}=\lvert G_{x}\rvert \mathrm{e}^{\mathrm{i}\phi_{x}}-\lvert G_{y}\rvert \mathrm{e}^{\mathrm{i}\phi_{y}},\\
B_{+}&=B_{-}=\sqrt{3}(\lvert G_{x}\rvert \mathrm{e}^{\mathrm{i}\phi_{x}}+\lvert G_{y}\rvert \mathrm{e}^{\mathrm{i}\phi_{y}}),\\
D_{1}&=0,\\
D_{2}&=0.
\end{align}
Thus, in this case, $w_\pm$ are independent of $\tilde{h}$ and the topological transition does not occur.

\section{Average square position $\Delta m^{2}$ in the OBC case} \label{app:delta_m2}
\begin{figure*}
    \centering
     \includegraphics[width=\linewidth,clip]{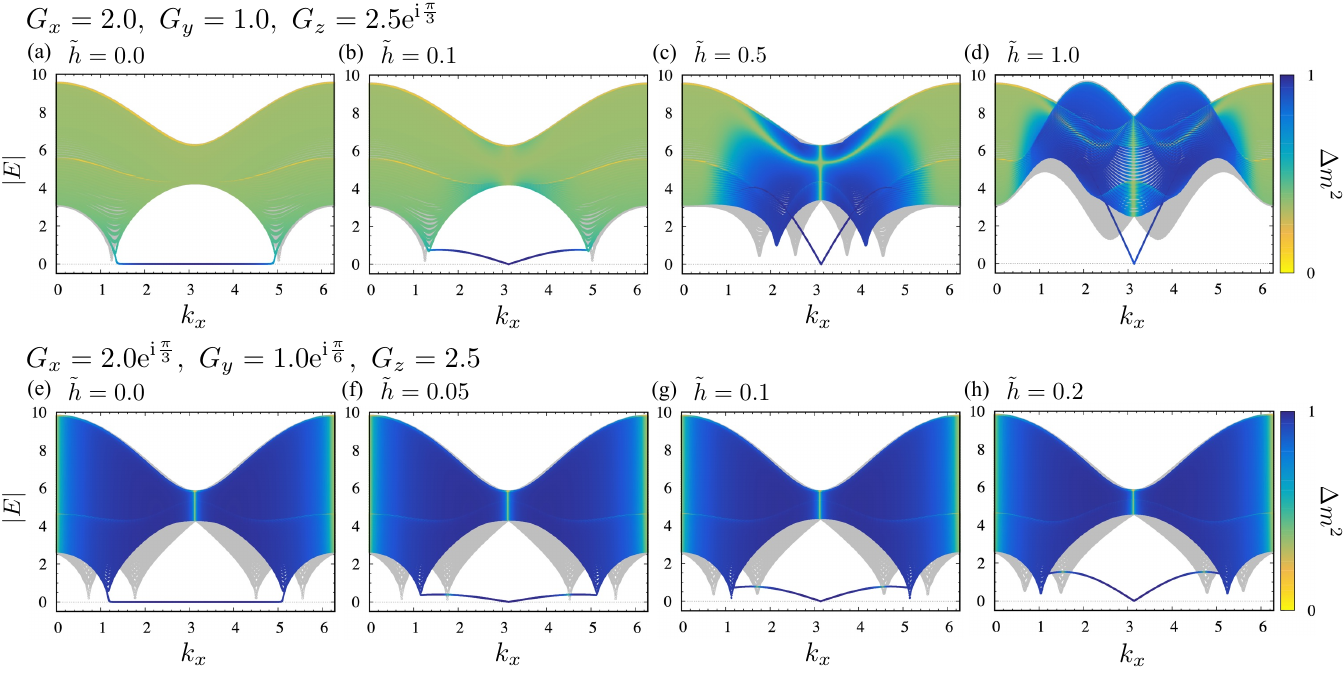}
    \caption{$k_{x}$ dependence of the absolute value of the complex energy eigenvalue with the open boundary condition at $G_{x}=2.0$, $G_{y}=1.0$, $G_{z}=2.5\mathrm{e}^{\mathrm{i}\frac{\pi}{3}}$, and (a) $\tilde{h}=0.0$, (b) $\tilde{h}=0.1$, (c) $\tilde{h}=0.5$, (d) $\tilde{h}=1.0$, and at $G_{x}=2.0\mathrm{e}^{\mathrm{i}\frac{\pi}{3}}$, $G_{y}=1.0\mathrm{e}^{\mathrm{i}\frac{\pi}{6}}$, $G_{z}=2.5$, and (e) $\tilde{h}=0.0$, (f) $\tilde{h}=0.05$, (g) $\tilde{h}=0.1$, and (h) $\tilde{h}=0.2$. 
    The color bar shows the average square position $\Delta m^{2}$.
    The data with the periodic boundary condition are shown in gray.
}
    \label{fig:ed_m2}
\end{figure*}

In this Appendix, we show supplemental data for the localization of the wave function in the skin effect in Sec.~\ref{subsec:skin_eff}. 
Figure~\ref{fig:ed_m2} shows the average square position corresponding to Figs.~\ref{fig:ed1} and \ref{fig:ed2}, which is defined as~\cite{Yang2021}
\begin{equation}
\Delta m^{2}(n, k_{x})=\frac{1}{M_0}\sum_{m=1}^{M}\left(m-M_0\right)^{2}\lvert \psi_{n}(k_{x}, m)\rvert^{2}, 
\end{equation}
where $M_0=\frac{M-1}{2}$. 
This quantity provide supplementary information to the average localization $\bar{m}$ in Eq.~\eqref{eq:mtilde};
it takes $1$ if the wave function is localized at the edges of the system, and $1/2$ if the wave function is spread over the entire system. 
Figures~\ref{fig:ed_m2}(a)--\ref{fig:ed_m2}(d), which correspond to Fig.~\ref{fig:ed1}, show that the isolated low-energy modes are localized at the edges of the system. Note that this is not evident solely from $\bar{m}$ for the case of $\tilde{h}=0$ in Fig.~\ref{fig:ed1}(a), since $\bar{m}$ takes the same value for the state equally localized at both edges and a state spread over the system. Meanwhile, in Figs.~\ref{fig:ed_m2}(f)--\ref{fig:ed_m2}(h), which corresponds to Fig.~\ref{fig:ed2}, $\Delta m^2$ also indicates that the low-energy modes are localized at the edges. Furthermore, it takes the value of $1/2$ at $k_{x}$ corresponding to the exceptional points of the PBC spectra. 
This supports the switching of the edge modes discussed in Sec.~\ref{subsec:skin_eff}.

\bibliography{library}

\end{document}